\documentclass[aps,prl,twocolumn,superscriptaddress]{revtex4-2}
\usepackage{tcolorbox}
\usepackage{xcolor}
\usepackage{amsmath,amssymb}
\usepackage{graphicx}
\usepackage{hyperref}
\usepackage{csquotes}
\usepackage{booktabs}
\usepackage{siunitx}
\usepackage{braket}
\usepackage{bm}
\usepackage{float} 
\begin{document}
	
	\title{Quantifying chirality of phonons}
	
	\author{Yu-Chi Huang}
    \email{y.huang5@tue.nl}
	\affiliation{Department of Physics, Institute of Science Tokyo, Tokyo 152-8551, Japan}
    \affiliation{Department of Applied Physics and Science Education, Eindhoven University of Technology, Eindhoven 5612 AP, Netherlands.}
	
	\author{Gakuto Kusuno}
	\affiliation{Department of Physics, Institute of Science Tokyo, Tokyo 152-8551, Japan}
	
	\author{Yusuke Hashimoto}
	\affiliation{Frontier Research Institute for Interdisciplinary Sciences, Tohoku University, Sendai 980-8578, Japan}

    \author{Dominik M. Juraschek}
    \affiliation{Department of Applied Physics and Science Education, Eindhoven University of Technology, Eindhoven 5612 AP, Netherlands.}
	
	\author{Hiroaki Kusunose}
	\affiliation{Department of Physics, Meiji University, Kawasaki 214-8571, Japan}
	
	\author{Takuya Satoh}
    \email{satoh@phys.sci.isct.ac.jp}
	\affiliation{Department of Physics, Institute of Science Tokyo, Tokyo 152-8551, Japan}
	\affiliation{Quantum Research Center for Chirality, Institute for Molecular Science, Okazaki 444-8585, Japan}

	\date{\today}
	
	\begin{abstract}
	Chiral phonons carry angular momentum with a handedness defined relative to their wave vector, as revealed by helicity-dependent optical phenomena, yet a unified mode-resolved and symmetry-weighted bulk characterization across crystals is lacking. We characterize each mode by the projection of its angular momentum onto the wave-vector direction and introduce two symmetry-weighted bulk measures, $G_0$ and $G_u$, of the net imbalance of thermally populated chiral phonons over the Brillouin zone. First-principles calculations for four chiral and four achiral crystals show that both bulk measures reverse sign between enantiomers and vanish in the achiral crystals studied, even when individual modes exhibit chirality. The bulk monopole $G_0$ takes opposite signs in right-handed Se and Te, demonstrating that its sign does not uniquely identify structural handedness. We also examine the relation between optical-phonon angular momentum and linear-in-wave-vector frequency splitting and compare the calculated splittings with circularly polarized Raman measurements in $\alpha$-quartz.
	\end{abstract}

	\maketitle
	
    Chiral phonons, lattice vibrations that carry angular momentum and possess a handedness, have attracted growing attention in condensed matter physics \cite{Juraschek2025, Sun2026Review, IshitoHgS, Ueda2023Nature}. Like circularly polarized light, they are a manifestation of dynamical chirality, in which the alignment of the angular momentum with the propagation direction changes sign under improper rotations. Dynamical chirality is distinct from, but connected to, conventional structural chirality, and it enters optical selection rules \cite{IshitoHgS, Zhang2015, Zhu2018, Chen2021ChiralPhonons, IshitoTe, Oishi2024, Zhang2022} and enantioselective physical processes \cite{Ohe2024, Feng2025}. Angular momentum is also transferred between spins, electrons, and lattice modes \cite{Dornes2019, Tauchert2022, Choi2024, Minakova2026, Nabei2026OrbitalSeebeck, Li2024, Hamada2018}.

	Existing approaches rest on the angular momentum of single modes tabulated across materials databases \cite{Wang2026}, on the helicity of a single soft mode at a displacive transition~\cite{GomezOrtiz2024}, or on a thermally averaged pseudoscalar for molecular chains \cite{Abraham2024}. A continuous symmetry measure has been used instead where the chirality originates at $\Gamma$ and no handedness can be assigned to a single mode~\cite{Fava2024PGO}. The inner product of phonon angular momentum with a polar vector that reduces to the wave vector was used to characterize the chirality of the phonons of tellurium, where thermally weighted Brillouin-zone sums of that product were also evaluated~\cite{IshitoTe}. The same inner product is the phonon helicity whose symmetry has been classified across the space groups \cite{Yang2026}, and a phononic analogue of optical zilch has since been proposed for helical crystals, with an explicit suggestion to use that inner product~\cite{TateishiKatoKishine2025}. The relation between such mode-level characterizations and symmetry-adapted bulk measures across three-dimensional crystals remains open.

    In this work, we combine mode-resolved and symmetry-weighted bulk measures of phonon chirality. The former characterizes each mode by the projection of its angular momentum onto the wave-vector direction, whereas the latter sums thermally weighted contributions over the Brillouin zone. We evaluate these measures for four chiral crystals ($\alpha$-quartz, Se, Te, and $\alpha$-HgS), a centrosymmetric achiral crystal (Si), and three noncentrosymmetric achiral crystals (GaAs, GaP, and ZnTe). We examine whether the bulk measures distinguish the chiral crystals from the achiral ones studied here and how their signs relate to structural handedness.
    
    \textit{Computational Details.} We calculate their phonon properties using density functional perturbation theory (DFPT) \cite{Baroni2001} as implemented in \textsc{abinit} \cite{ABINIT_2020}, and evaluate both measures from the eigenfrequencies and eigendisplacements obtained by diagonalizing the dynamical matrix, the Fourier transform of the interatomic force constants. The computational parameters are given in Supplemental Material I~\cite{supple}.
    
    \textit{Wave-Vector-Resolved Dynamical Chirality.} Following the concept of true chirality~\cite{Barron1986, Barron1986_2}, we characterize a phonon mode $j$ at wave vector $\mathbf{k}$ by the projection of its angular momentum $\mathbf{L}_j(\mathbf{k})$~\cite{Juraschek2025, IshitoHgS, Ueda2023Nature, Zhang2014} onto the wave-vector direction. Because angular momentum is an axial vector whereas wave vector is a polar vector, their scalar product changes sign under mirror reflection, and because both are time-odd, it remains invariant under time reversal. This time-even pseudoscalar motivates the wave-vector-resolved measure $\mathbf{L}_j(\mathbf{k})\cdot\hat{\mathbf{k}}$ for $\mathbf{k}\neq0$. Its sign specifies the handedness of the phonon mode, while its magnitude gives the size of the angular-momentum component along the propagation direction (Fig.~1). Here $\hat{\mathbf{k}}$ denotes the normalized wave vector. The values then lie within $[-\hbar, \hbar]$, which enables a clear visualization of mode-resolved chirality across the Brillouin zone. 
	
	\begin{figure}
		\makebox[\textwidth][l]{
	    \includegraphics[width=0.5\textwidth]{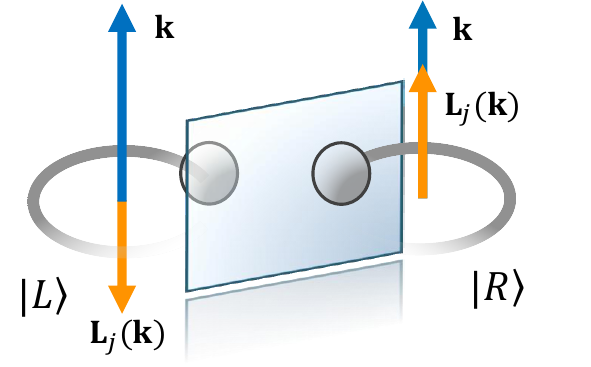}
		}
		\caption{A chiral phonon in a branch $j$ carries collinear angular momentum, $\mathbf{L}_j(\mathbf{k})$, and wave vector, $\mathbf{k}$, in the two states labelled $|L\rangle$ and $|R\rangle$. A reflection reverses the sense of rotation, while preserving the direction of motion of the phonon wave packet, switching between parallel and antiparallel alignments of $\mathbf{L}_j(\mathbf{k})$ and $\mathbf{k}$. The collinear arrangement is drawn for illustration, and the two dynamical chiral enantiomers are distinguished in general by the sign of $\mathbf{L}_j(\mathbf{k})\cdot\hat{\mathbf{k}}$.}
	\end{figure}

    \begin{figure*}[t]
		\makebox[\textwidth][c]{
	    \includegraphics[width=0.9\textwidth]{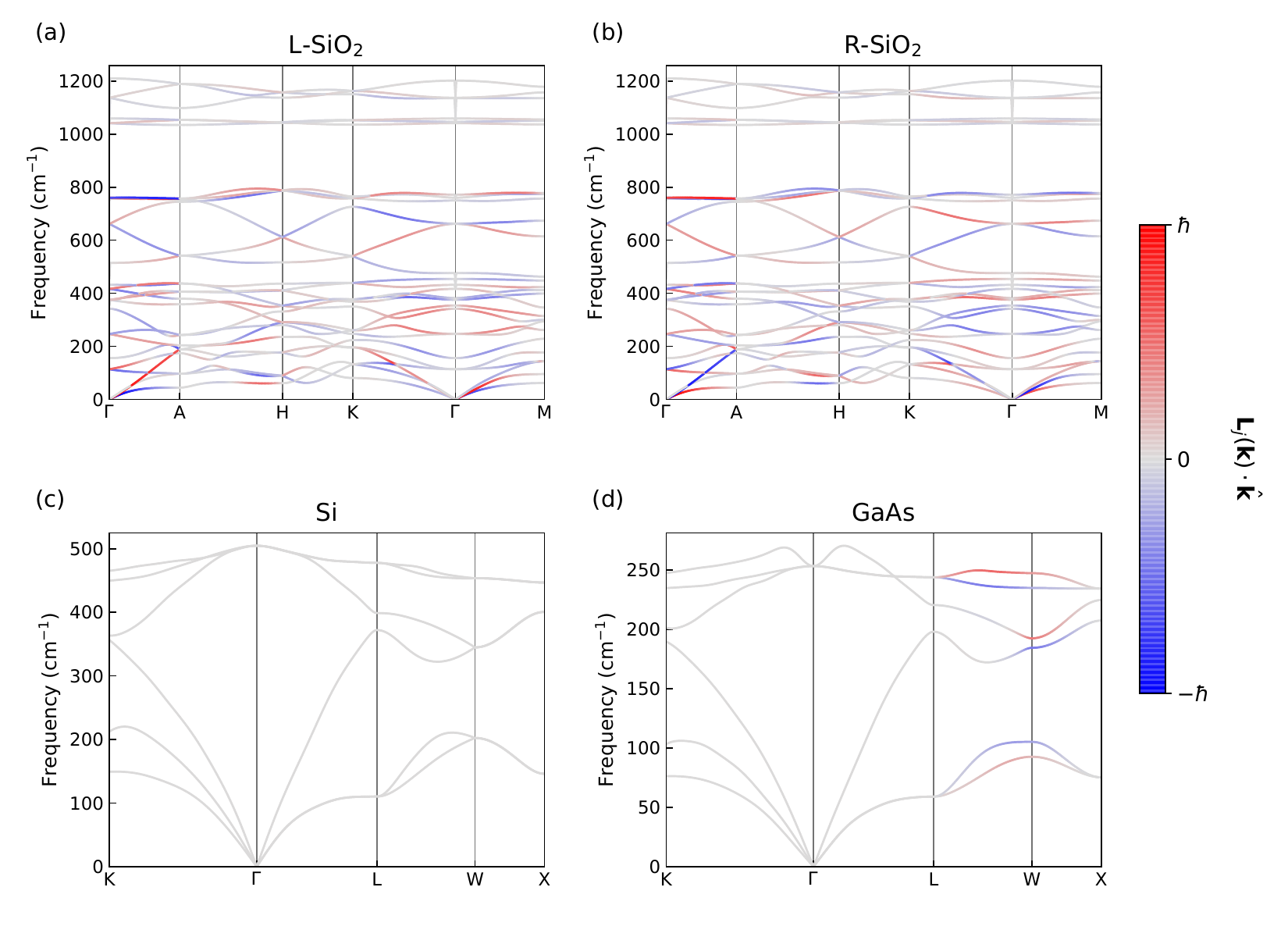}
		}
		\caption{Phonon dispersions colored by the wave-vector-resolved dynamical chirality $\mathbf{L}_j(\mathbf{k}) \cdot \hat{\mathbf{k}}$ in (a) L-SiO$_2$, (b) R-SiO$_2$, (c) Si, and (d) GaAs along high-symmetry paths in reciprocal space. The sign of the wave-vector-resolved dynamical chirality flips between the two enantiomers of $\alpha$-quartz and is spread broadly along these paths. Si is centrosymmetric-achiral and shows no phonon chirality throughout the band structure \cite{Coh2023}. GaAs is noncentrosymmetric-achiral and shows phonon chirality along specific high-symmetry paths.}
        \label{fig:comparison1}
	\end{figure*}

    Figure~\ref{fig:comparison1} shows the wave-vector-resolved dynamical chirality mapped onto the phonon band structures of three representative materials: $\alpha$-quartz (chiral, $D_3$), Si (centrosymmetric achiral, $O_h$), and GaAs (noncentrosymmetric achiral, $T_d$). Results for the remaining five materials are given in Supplemental Material II~\cite{supple}.

    The two enantiomers of $\alpha$-quartz carry opposite signs along the high-symmetry paths shown [Figs.~2(a) and 2(b)], which is a reversal of the handedness of the chiral phonon modes. In Si, the wave-vector-resolved dynamical chirality vanishes within numerical accuracy [Fig.~2(c)], whereas in GaAs it takes finite values in the vicinity of the W point [Fig.~2(d)]. Chiral phonon modes therefore occur without structural chirality.

	\textit{Bulk Dynamical Chirality.} The wave-vector-resolved measure does not directly provide a bulk quantity. At fixed $\mathbf{k}$ the complete branch sum of phonon angular momentum vanishes by the sum rule, so a structure factor common to all branches cannot by itself prevent that cancellation. The branch-dependent phonon occupation factor can prevent it, while a periodic structure factor provides the symmetry-adapted weight. Using the lowest-order symmetry-adapted structure factor $\mathbf{F}_1(\mathbf{k})$, we define the electric toroidal monopole $G_0$, the isotropic component of the bulk true chirality, and the electric toroidal quadrupole $G_u$, its uniaxial anisotropy \cite{IshitoTe, Kusunose2024, Kishine2022, Inda2024, Oiwa2022, Hayami2024, Tsunetsugu2023}, as
    \begin{align}
    G_0(T) &= \frac{1}{N_0 \hbar}\sum_{j} \sum_{\mathbf{k}}
    f(\omega_j(\mathbf{k}))
    \left[\mathbf{L}_j(\mathbf{k}) \cdot \mathbf{F}_1(\mathbf{k}) \right], \nonumber\\
    G_u(T) &= \frac{1}{N_0 \hbar}\sum_{j} \sum_{\mathbf{k}}
    f(\omega_j(\mathbf{k}))
    \frac{1}{2}\big[3L^z_j(\mathbf{k}) F^z_1(\mathbf{k})\label{eq:bulk}
    \\& \hspace{4cm} - \mathbf{L}_j(\mathbf{k}) \cdot \mathbf{F}_1(\mathbf{k})\big].\nonumber
    \end{align}
    
    \noindent{}Here, $N_0$ is the number of lattice points, $j$ labels the phonon branch, and $f(\omega_j(\mathbf{k}))=n_\text{B}(\omega_j(\mathbf{k}))+1/2$, where $n_\text{B}$ is the Bose-Einstein distribution. The zero-point term cancels in the complete phonon-branch sum (Supplemental Material III~\cite{supple}). We evaluate $G_0$ and $G_u$ at $T=300$~K. $\mathbf{F}_1(\mathbf{k})$ is a symmetry-adapted periodic structure factor~\cite{IshitoTe, Hayami2024}. Its explicit forms are given in Supplemental Material IV~\cite{supple}. Building on the tellurium construction of Ref.~\cite{IshitoTe}, we fix the normalization of the structure factor and the convention for $G_u$, and apply them to the eight crystals studied here. Both $\mathbf{L}_j(\mathbf{k})$ and $\mathbf{F}_1(\mathbf{k})$ are odd in $\mathbf{k}$, so their scalar product is even and need not cancel in the Brillouin-zone sum, even though the total phonon angular momentum itself vanishes under time-reversal symmetry~\cite{Nabei2026OrbitalSeebeck}.
    Table~\ref{tab:materials} summarizes the calculated $G_0$ and $G_u$ for eight crystals. For the four chiral materials, both $G_0$ and $G_u$ take finite values and show sign reversal between opposite enantiomers. In contrast, all four achiral materials exhibit vanishing values of $G_0$ and $G_u$ within numerical accuracy, despite finite wave-vector-resolved chirality at individual wave vectors. This cancellation reflects the symmetry of the system~\cite{Yang2026}, indicating that $G_0$ and $G_u$ capture the net imbalance of chiral phonons. Notably, although $G_0$ and $G_u$ change sign between opposite enantiomers, their absolute sign depends on the population of left- and right-handed phonons and thus does not uniquely specify the handedness of the crystal~\cite{Harris1999}.

    \textit{Material dependence of bulk dynamical chirality.} Table~I
    shows that R-Se and R-Te have opposite values of $G_0$, $+0.10$ and
    $-0.08$, despite having the same structural handedness. Both belong to the point group $D_3$, whose symmetry classification of phonon helicity~\cite{Yang2026} does not fix the sign of $G_0$ for a given enantiomer, so this difference is a result of the calculation rather than of symmetry. To understand
    this difference, we resolve Eq.~(\ref{eq:bulk}) into its $c$-axis
    and basal-plane
    contributions, $G_{zz}=(G_0+2G_u)/3$ and
    $G_{xx}+G_{yy}=2(G_0-G_u)/3$~\cite{Hayami2024} in the normalization of Eq.~(\ref{eq:bulk}), evaluated from the unrounded values. These two contributions sum to $G_0$. The $c$-axis contributions are
    similar for R-Se and R-Te ($-0.16$ and $-0.15$), whereas their basal-plane contributions differ substantially ($+0.26$ and $+0.06$). The stronger positive basal-plane contribution makes $G_0$ positive in R-Se, whereas the negative $c$-axis contribution dominates in R-Te. The $G_0$ of R-Se results from a partial
    cancellation of contributions with opposite signs and is therefore the
    least robust entry in Table~I\@. Its sign is sensitive to the
    normalization of the $c$-axis structure-factor component and to the
    supercell length used in the finite-displacement check (Supplemental Material IV and V).

    For comparison, the positive $G_0=+0.31$ of R-SiO$_2$ is
    dominated by the basal-plane contribution ($+0.28$), with only $+0.03$
    from the $c$-axis component. In R-HgS, both contributions are
    positive, and the large $G_0=+0.65$ is dominated by the basal-plane contribution ($+0.62$).

	\begin{table}
        \caption{Calculated values of the bulk dynamical chirality as characterized by $G_0$ and $G_u$ for selected chiral and achiral materials at $T=300$~K. L and R denote left- and right-handed enantiomorphic structures, respectively. Values given as $0.00$ for the achiral materials vanish within numerical accuracy.}
        \centering
        \begin{tabular}{lcc@{\hspace{20pt}}lcc}
        \toprule
        \textbf{Crystal} & $G_0$ & $G_u$ & \textbf{Crystal} & $G_0$ & $G_u$ \\
        \midrule
        \multicolumn{3}{c}{Chiral materials} & \multicolumn{3}{c}{Achiral materials} \\
        L-SiO$_2$ & $-0.31$ & $0.11$ & Si   & $0.00$ & $0.00$ \\
        R-SiO$_2$ & $0.31$ & $-0.11$ & GaAs & $0.00$ & $0.00$ \\
        L-Se      & $-0.10$ & $0.29$ & GaP  & $0.00$ & $0.00$ \\
        R-Se      & $0.10$  & $-0.29$ & ZnTe & $0.00$ & $0.00$ \\
        L-Te      & $0.08$  & $0.18$ &       &        &        \\
        R-Te      & $-0.08$ & $-0.18$ &       &        &        \\
        L-HgS     & $-0.65$ & $0.29$ &       &        &        \\
        R-HgS     & $0.65$  & $-0.29$ &       &        &        \\
        \bottomrule
        \end{tabular}
        \vspace{2pt}
        \begin{flushleft}
        \footnotesize
        Chiral materials: space groups $\mathrm{P}3_221$ (L) and $\mathrm{P}3_121$ (R) (point group $D_3$).
        Achiral materials: $\mathrm{Fd}\bar{3}\mathrm{m}$ ($O_h$) for Si, and $\mathrm{F}\bar{4}3\mathrm{m}$ ($T_d$) for GaAs, GaP, and ZnTe.
        The $q$-meshes, pseudopotentials, and convergence tests are given in Supplemental Material~I, and the frame convention in Supplemental Material~IV.
        \end{flushleft}
        \label{tab:materials}
    \end{table}

    \begin{figure*}[t]
	\centering
	    \includegraphics[width=0.90\textwidth]{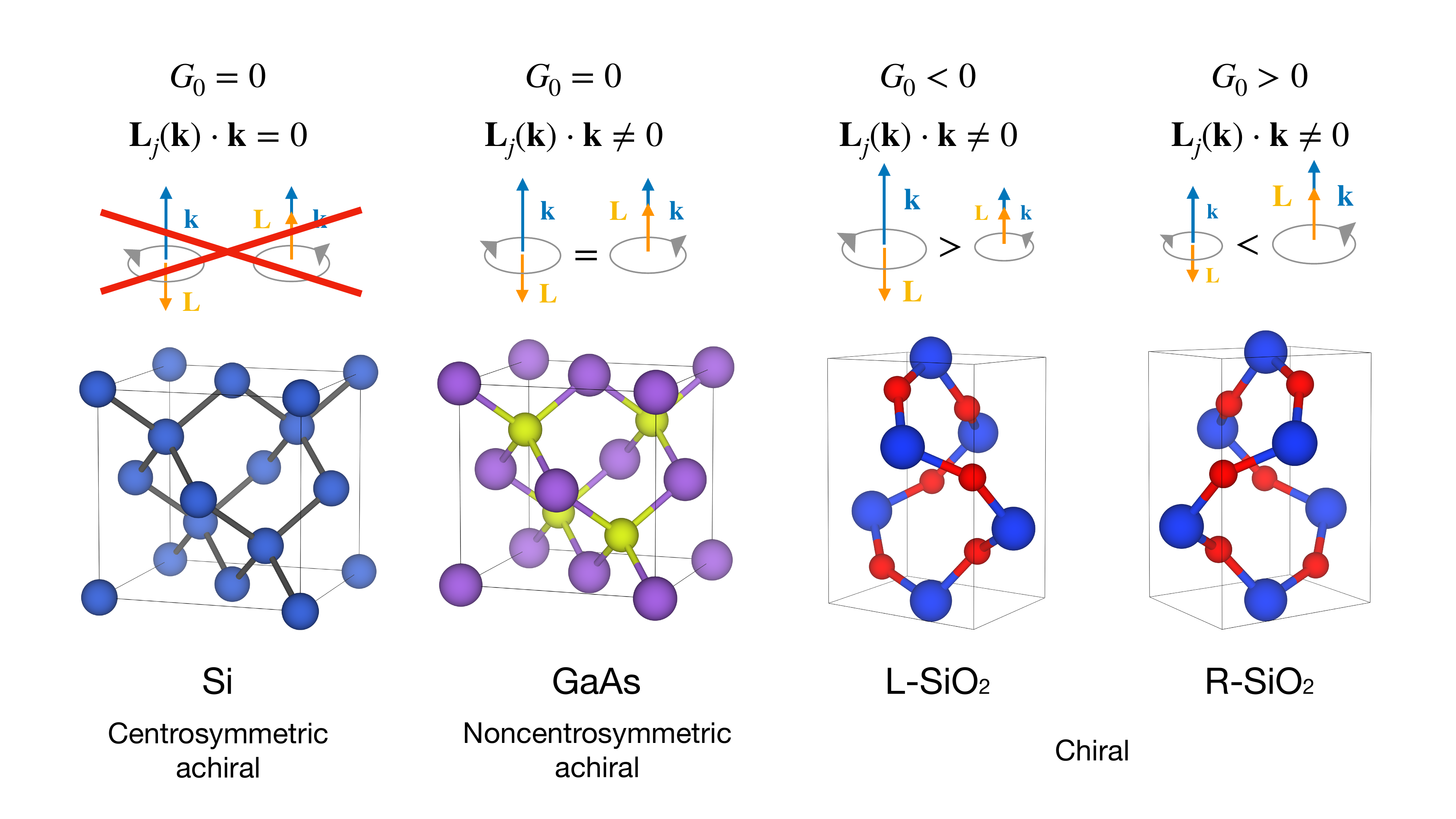}
	\caption{Schematic comparison of wave-vector-resolved and bulk dynamical chirality in centrosymmetric Si, noncentrosymmetric achiral GaAs, and the two enantiomers of $\alpha$-quartz. The wave-vector-resolved quantity $\mathbf{L}_j(\mathbf{k})\cdot\hat{\mathbf{k}}$ characterizes individual modes, whereas $G_0$ measures a symmetry-weighted bulk imbalance. Local chirality can be finite even when $G_0=0$, whereas enantiomorphic crystals show opposite signs of the bulk dynamical chirality.}
        \label{fig:comparison2}
	\end{figure*}
    
    \textit{Linear-in-$k$ splitting.}
    We next examine how the dynamical chirality of individual
    phonon modes can enter their dispersion~\cite{Xie2026}. Schematically, we consider a
    contribution of the form
    \[
        \omega_j(\mathbf{k})=
        \omega_j^{(0)}(\mathbf{k})
        +\gamma_j\,\mathbf{L}_j(\mathbf{k})\cdot\mathbf{k}
        +\cdots,
    \]
    where $\omega_j^{(0)}$ denotes a reference dispersion and
    $\gamma_j$ characterizes the coefficient of the chirality-dependent
    term. This symmetry-motivated form carries the pseudoscalar, electric
    toroidal monopole ($G_0$-type) structure of the
    wave-vector-resolved dynamical chirality defined above. For the
    $D_3$ crystals considered here, symmetry also permits a uniaxially
    anisotropic bilinear term of electric toroidal quadrupole
    ($G_u$-type) symmetry, proportional to
    $[3L^z_j(\mathbf{k})k_z-\mathbf{L}_j(\mathbf{k})\cdot\mathbf{k}]/2$.
    For $\mathbf{k}\parallel c$, the two reduce to the same
    $L^z_jk_z$ and cannot be distinguished in the optical $E$-doublet
    splitting, so any such contribution is absorbed into the
    mode-specific effective coefficient $\gamma_E$ discussed below.
    This decomposition is not unique for arbitrary phonon branches.
    An observable example is the splitting near $\Gamma$ of an optical $E$ doublet, which can be probed by circularly polarized Raman spectroscopy~\cite{IshitoHgS, PineDresselhaus1969}. We label the optical $E$ doublets of each crystal $E^{(1)},E^{(2)},\ldots$
    in order of increasing frequency at $\Gamma$ (Supplemental
    Material~VI). For $\mathbf{k}\parallel c$ the threefold screw symmetry
    distinguishes the two branches of a doublet by crystal angular momentum
    (also called pseudoangular momentum) $m=\pm1$ and allows a linear
    splitting~\cite{PineDresselhaus1969, Zhang2022, IshitoHgS},
    \begin{align}
        \omega_+-\omega_-&=2Bk_z+O(k_z^3).
        \label{eq:split}
    \end{align}
    Here, $B$ is half the linear slope of the frequency difference. The
    symmetry derivation, quadratic dispersion, and acoustic case are
    discussed in Supplemental Material~VI\@.

    For an optical $E$ doublet, we define the reference dispersion as
    the average of the two phonon frequencies,
    $\omega_E^{(0)}(k_z)=[\omega_+(k_z)+\omega_-(k_z)]/2$.
    For the lowest such doublet of each crystal, the two states at $\Gamma$
    carry opposite mechanical angular momenta of equal magnitude,
    $|\langle L_z\rangle_\pm(0)|=\hbar\ell_0$, where $\ell_0$ denotes their
    magnitude in units of $\hbar$ (Supplemental Material~VI). The atoms
    occupy general and $C_2$ Wyckoff positions rather than $C_3$-invariant ones,
    so $\ell_0$ is not pinned to unity~\cite{ZhangS2026}.
    For $\ell_0\neq0$ the dynamical matrix of the doublet acquires the term
    $\gamma_Ek_zL_z/\hbar$ at linear order, with $|\gamma_E|=|b_1|/\ell_0$.
    It reduces to the coefficient $\gamma_\sigma$ of
    Ref.~\cite{Tsunetsugu2026} when $\ell_0=1$.
    In the chiral $D_3$ crystals studied here, the absence of mirror
    symmetry allows the two rotational branches of an optical $E$ doublet to
    acquire opposite frequency shifts linear in $k_z$, with time-reversal
    symmetry relating $\omega_+(k_z)$ to $\omega_-(-k_z)$. Their opposite
    mechanical angular momenta at $\Gamma$ allow this splitting to be expressed
    in terms of $L_zk_z$, making the connection between lattice rotation and
    dispersion explicit. The coefficient $\gamma_E$ characterizes this relation
    for each doublet without implying a universal microscopic coupling.

    In Te, the $E^{(1)}$ and $E^{(2)}$ doublets have markedly different mechanical angular momenta ($\ell_0=0.9$ and $0.1$), whereas direct finite-$\mathbf{q}$ DFPT gives comparable splitting slopes ($|B|=31.7$ and $42.5$~cm$^{-1}$\,\AA, respectively). Table~\ref{tab:splitting} lists the interpolated slopes, and the comparison between the two methods is given in Supplemental Material~VI.

    Table~\ref{tab:splitting} summarizes the $\Gamma$-point frequency
    $\omega_0=\omega_\pm(0)$, the slope $|B|$, and the
    mechanical angular momentum $\ell_0$ for the lowest optical $E$ doublet
    of each material. For comparison with
    Refs.~\cite{Tsunetsugu2023, Tsunetsugu2026}, we also give
    $|b_1|=2\omega_0|B|$, where $b_1$ is defined by the splitting of the
    dynamical-matrix eigenvalues, $\omega_+^2-\omega_-^2=2b_1k_z+O(k_z^3)$.
    It corresponds to the splitting coefficient $\Gamma_\alpha$ of
    Ref.~\cite{Tsunetsugu2023}, where $\alpha$ labels the branch, and $B$ to
    the coefficient $c_\sigma$ of Ref.~\cite{Tsunetsugu2026}.
    Among the four materials, $|B|$ of the lowest doublet is largest for $\alpha$-SiO$_2$
    ($77.9$~cm$^{-1}$\,\AA) and smallest for $\alpha$-HgS
    ($22.6$~cm$^{-1}$\,\AA), while their $|b_1|$ values differ by a factor of
    nine because of the additional frequency factor $\omega_0$.

    Interestingly, Table~\ref{tab:materials} shows the reverse for the two
    extremes: $|G_0|=0.31$ for $\alpha$-SiO$_2$
    and $0.65$ for $\alpha$-HgS\@. This contrast reflects the different
    information contained in the two quantities. Whereas $B$ characterizes
    the dispersion of a single optical doublet near $\Gamma$, $G_0$ sums over
    all phonon branches throughout the Brillouin zone, with thermal and
    symmetry-dependent weights that allow contributions of opposite signs
    to cancel.

    These mode-specific splittings can be compared with circularly
    polarized Raman measurements in $\alpha$-quartz. In backscattering
    geometry, the phonon wave vector along the $c$ axis is
    $k_z\simeq4\pi n/\lambda$, where $n$ is the refractive index and
    $\lambda$ is the vacuum wavelength. In the geometry of
    Ref.~\cite{Oishi2024} the circular polarizations distinguish the two
    branches of opposite crystal angular momentum~\cite{IshitoHgS, ZhangS2026, Watanabe2025ROA}. For $\alpha$-quartz at
    $\lambda=532$~nm the calculated splittings of the eight optical $E$
    doublets range from $0.06$ to $1.74$~cm$^{-1}$. The largest occurs for
    $E^{(5)}$ at $661$~cm$^{-1}$ ($1.74$~cm$^{-1}$), tentatively assigned to the measured
    peak at $695$~cm$^{-1}$ ($1.53$~cm$^{-1}$). For $E^{(1)}$ the calculated splitting is
    $0.57$~cm$^{-1}$ at $114$~cm$^{-1}$, compared with $0.31$~cm$^{-1}$ at the
    measured $128$~cm$^{-1}$ peak~\cite{Oishi2024}. The calculations
    reproduce the order of magnitude of the observed splittings.
    The reversal of the polarization-resolved frequency ordering
    between the two enantiomers of quartz~\cite{Oishi2024} provides a
    spectroscopic signature of the handedness-dependent linear splitting
    $2Bk_z$. Because backscattering selects a small but finite $k_z$,
    the peak separation probes the mode-specific slope $B$.
    This spectroscopic information is distinct from the mechanical angular
    momentum of the phonon and from the bulk dynamical chirality
    characterized by $G_0$ and $G_u$. Quantitative differences between
    calculation and experiment are detailed in Supplemental Material~VI.
	\begin{table}
        \caption{Linear-in-$k$ splitting of $E^{(1)}$, the lowest
        optical $E$ doublet of each crystal, for $\mathbf{k}\parallel c$, from the
        same dynamical matrices as Table~\ref{tab:materials}. Magnitudes are
        given, and $B$ and $b_1$ reverse sign between the two
        enantiomers. The slopes are interpolated along $\mathbf{k}\parallel c$,
        and Supplemental Material~VI compares them with direct
        calculations at finite $\mathbf{q}$ for Te and Se.}
        \centering
        \small
        \begin{tabular}{lcccc}
        \toprule
        & $\omega_0$ & $|B|$ & $|b_1|$ & $\ell_0$ \\
        \textbf{Crystal} & (cm$^{-1}$) & (cm$^{-1}$\,\AA)
            & (cm$^{-2}$\,\AA) & \\
        \midrule
        $\alpha$-SiO$_2$ & $113.7$ & $77.9$ & $1.8\times10^{4}$ & $0.54$ \\
        Se              & $127.8$ & $63.7$
            & $1.6\times10^{4}$ & $0.81$ \\
        Te              & $81.1$  & $32.7$
            & $5.3\times10^{3}$ & $0.90$ \\
        $\alpha$-HgS    & $43.4$ & $22.6$ & $2.0\times10^{3}$ & $0.89$ \\
        \bottomrule
        \end{tabular}
        \label{tab:splitting}
    \end{table}

    \textit{Discussion.} Figure~\ref{fig:comparison2} summarizes the
    distinction between the wave-vector-resolved and bulk measures. In GaAs,
    local phonon chirality is possible without structural chirality, as a
    lattice-dynamical effect allowed whenever inversion is broken and time
    reversal preserved~\cite{Coh2023, Ueda2025}, but its
    symmetry-weighted contributions cancel in the Brillouin-zone sum. In the enantiomorphic crystals studied here, by contrast, the bulk measures reverse sign with structural handedness. Finite-displacement calculations using \textsc{vasp}
    and \textsc{phonopy} provide a cross-check of the calculated
    values of $G_0$ and $G_u$ for the right-handed chiral crystals,
    reproducing the sign of both measures for all four and agreeing with the
    DFPT values to within $0.05$ in $G_0$ and $0.08$ in $G_u$.
    The same approach reproduces
    the symmetry-required cancellation in GaAs, GaP, and ZnTe
    (Supplemental Material V~\cite{supple}).

    Whether the bulk dynamical chirality is an order parameter for
    chirality is a natural question~\cite{Bousquet2025, Fava2025,
    Spaldin2026, GomezOrtiz2026}. $G_0$ and $G_u$ are constrained by crystal
    symmetry, vanish in high-symmetry phases, and reverse sign between
    enantiomers, but whether either can serve as a thermodynamic order
    parameter with a conjugate field remains to be established~\cite{Bousquet2025, GomezOrtiz2026}. Here we use them as
    dynamical measures of the symmetry-weighted imbalance of thermally
    populated chiral phonons, which purely structural descriptors do not
    capture~\cite{Khan2025CCM}.

    A further connection is with transport. The $c$-axis component
    $G_{zz}$ has the same symmetry properties~\cite{Hayami2024, Yang2026} as the phonon
    Edelstein response coefficient $\alpha_{zz}$, which describes phonon
    angular momentum induced by a temperature
    gradient~\cite{Hamada2018}, a response that has since been
    measured in Te~\cite{ZhangH2025}. Both are built from the same
    helicity kernel, but the transport response weights it by the group
    velocity whereas $G_{zz}$ does not,
    so the two sums are not controlled by the same doublets, and their symmetry correspondence does not by itself establish a quantitative relation between them. Whether
    $G_{zz}$ can quantitatively predict the phonon Edelstein response
    therefore remains to be established. A second-order effective electron--phonon coupling to phonon mechanical angular momentum has also been derived for a helical model~\cite{KatoYokoshiKishine2026}. Identifying fields or driving schemes that couple
    selectively to chiral phonons, and responses tied to the bulk measures,
    remains for future work. The formulation extends to other chiral point
    groups.

    \textit{Summary.} We have introduced complementary wave-vector-resolved
    and bulk measures of phonon dynamical chirality. The bulk measures reverse sign between
    enantiomers, but their magnitudes and absolute signs are
    determined by collective phonon dynamics and do not provide
    universal labels of structural handedness. Circularly polarized
    Raman spectroscopy provides complementary information on the linear-in-$k$
    splitting of individual optical doublets rather than directly measuring
    these bulk quantities. This framework enables quantitative comparisons
    of dynamical chirality across crystals.
	
	\textit{Acknowledgement.} This work was supported in part by JSPS KAKENHI (Grant Nos. \mbox{JP21H01032}, \mbox{JP22H01154}, \mbox{JP23K03288}, and \mbox{JP26H02234}), MEXT X-NICS (Grant No. \mbox{JPJ011438}), NINS OML Project (Grant No. \mbox{OML012301}), JST CREST (Grant No. \mbox{JPMJCR24R5}), and JST ERATO (Grant No. \mbox{JPMJER2503}). D.M.J. acknowledges support from the ERC Starting Grant CHIRALPHONONICS, no. \mbox{101166037}. The authors thank SAKURA internet Inc. for providing the GPU computing resources used in this research.

    \bibliography{chiral_phonons.bib,SM.bib}

\end{document}


\title{Supplemental Material:\\Quantifying chirality of phonons}
	
	\author{Yu-Chi Huang}
	\affiliation{Department of Physics, Institute of Science Tokyo, Tokyo 152-8551, Japan}
	\affiliation{Department of Applied Physics and Science Education, Eindhoven University of Technology, Eindhoven 5612 AP, Netherlands.}
	
	\author{Gakuto Kusuno}
	\affiliation{Department of Physics, Institute of Science Tokyo, Tokyo 152-8551, Japan}

	\author{Yusuke Hashimoto}
	\affiliation{Frontier Research Institute for Interdisciplinary Sciences, Tohoku University, Sendai 980-8578, Japan}	
		
	\author{Dominik M. Juraschek}
	\affiliation{Department of Applied Physics and Science Education, Eindhoven University of Technology, Eindhoven 5612 AP, Netherlands.}
	
	\author{Hiroaki Kusunose}
	\affiliation{Department of Physics, Meiji University, Kawasaki 214-8571, Japan}
	
	\author{Takuya Satoh}
	\affiliation{Department of Physics, Institute of Science Tokyo, Tokyo 152-8551, Japan}
	\affiliation{Quantum Research Center for Chirality, Institute for Molecular Science, Okazaki 444-8585, Japan}
	
	\date{\today}
	
	\maketitle

    \renewcommand{\thefigure}{S\arabic{figure}}
    \setcounter{figure}{0}
    \renewcommand{\theequation}{S\arabic{equation}}
    \setcounter{equation}{0}
	\renewcommand{\thetable}{S\arabic{table}}
    \setcounter{table}{0}
	
	\section{Computational Parameters}
\label{sec:comp_params}
	
	Table~\ref{tab:params} summarizes the computational parameters for the eight materials discussed in the main text.
	Phonon properties were computed using density functional perturbation theory (DFPT)~\cite{Baroni2001} as implemented in the \textsc{abinit}~code~\cite{ABINIT_2020, Verstraete2025}. The dynamical matrices were computed with version~10, and the interatomic force constants and the Brillouin-zone sums with \textsc{anaddb}~9.10.4.
	In addition to the DFPT calculations, finite-displacement calculations with \textsc{vasp}~\cite{Kresse1996} were performed for all eight materials: the four chiral crystals ($\alpha$-SiO$_2$, Se, Te, and $\alpha$-HgS) and the four achiral crystals (Si, GaAs, GaP, and ZnTe). The corresponding phonon-analysis procedure uses \textsc{phonopy}~\cite{Togo2015}, as described in Sec.~\ref{sec:vasp}.
	The plane-wave cutoff energies and $k$-point meshes were chosen to ensure 	convergence of both total energies and phonon frequencies. Spin-orbit coupling was not included.
	GGA-PBE denotes the Perdew--Burke--Ernzerhof generalized gradient approximation~\cite{Perdew1996}.

	Seven structural cells, those of $\alpha$-SiO$_2$, Se, Te, and the four achiral crystals, were relaxed within PBE with \textsc{abinit}~9.10.4 by releasing both the cell and the ions, with a force threshold of $10^{-5}$~Ha/bohr. For the achiral crystals this gives $a=5.4701$ (Si), $5.7467$ (GaAs), $5.5087$ (GaP), and $6.1719$~\AA\ (ZnTe), the only free parameter being the lattice constant. For $\alpha$-SiO$_2$, Se, and Te the relaxation used the same pseudopotentials, plane-wave cutoff, and $k$-mesh as the DFPT phonon calculation.

	For $\alpha$-HgS we use instead a cell relaxed with PBEsol~\cite{Perdew2008}, that of the phonon database entry of Ref.~\cite{TogoPhonondb}. Both the DFPT and the finite-displacement calculations use GGA-PBE on that cell without relaxing it again.

	The pseudopotentials used in the DFPT calculations are norm-conserving and were generated with \textsc{oncvpsp}~3.2.3.1~\cite{Hamann2013}. Si, O, Se, and Te are byte-identical to the PseudoDojo v0.4 PBE table~\cite{vanSetten2018}, and the Ga, As, P, and Zn files are its entries Ga-d, As-d, P, and Zn-sp, identified by the generator, the core radii, the valence charge, and the generation date. The Hg and S files come with the input files of Ref.~\cite{IshitoHgS} and share the same generator and release.

	Table~\ref{tab:struct_psp} lists the structural parameters and valence-electron counts of the four chiral crystals.
	In the DFPT calculations, both enantiomers of all four chiral crystals were computed explicitly, and they reproduce the sign reversal of $G_0$ and $G_u$.
	
	\begin{table*}[htbp]
		\centering
		\small
		\caption{Computational parameters for all investigated materials, given in parallel for the DFPT calculations with \textsc{abinit} and the finite-displacement (FD) calculations with \textsc{vasp} and \textsc{phonopy}. The phonon sampling column lists the $q$-mesh for DFPT and the supercell for FD, the supercell being the largest of the series used for that crystal. NAC denotes the non-analytic correction. For Si it is immaterial in the DFPT calculation because the Born effective charges vanish by symmetry. For the chiral crystals the FD entries refer to the right-handed structures.}
		\begin{tabular}{llcccccc}
			\toprule
			Material & Space group & Method & Cutoff & $k$-mesh & Phonon sampling & XC & NAC \\
			\midrule
			\multicolumn{8}{l}{(Chiral)} \\
			\multirow{2}{*}{$\alpha$-SiO$_2$ (L/R)} & \multirow{2}{*}{$\text{P}3_{2}21/\text{P}3_{1}21$} & DFPT & 60 Ha & $6\times6\times6$ & $8\times8\times6$ & GGA-PBE & Yes \\
			 &  & FD & 520 eV & $2\times2\times2$ & $4\times4\times3$ & GGA-PBE & Yes \\
			\multirow{2}{*}{Se (L/R)} & \multirow{2}{*}{$\text{P}3_{2}21/\text{P}3_{1}21$} & DFPT & 40 Ha & $8\times8\times6$ & $14\times14\times6$ & GGA-PBE & Yes \\
			 &  & FD & 408 eV & $1\times1\times2$ & $5\times5\times6$ & GGA-PBE & Yes \\
			\multirow{2}{*}{Te (L/R)} & \multirow{2}{*}{$\text{P}3_{2}21/\text{P}3_{1}21$} & DFPT & 40 Ha & $8\times8\times6$ & $14\times14\times6$ & GGA-PBE & Yes \\
			 &  & FD & 408 eV & $2\times2\times2$ & $5\times5\times4$ & GGA-PBE & Yes \\
			\multirow{2}{*}{$\alpha$-HgS (L/R)} & \multirow{2}{*}{$\text{P}3_{2}21/\text{P}3_{1}21$} & DFPT & 60 Ha & $6\times6\times4$ & $12\times12\times6$ & GGA-PBE & Yes \\
			 &  & FD & 600 eV & $\Gamma$ & $4\times4\times4$ & GGA-PBE & Yes \\
			\midrule
			\multicolumn{8}{l}{(Achiral)} \\
			\multirow{2}{*}{Si} & \multirow{2}{*}{$\text{Fd}\bar{3}\textit{m}$} & DFPT & 60 Ha & $6\times6\times6$ & $10\times10\times10$ & GGA-PBE & --- \\
			 &  & FD & 408 eV & $\Gamma$ & $4\times4\times4$ & GGA-PBE & No \\
			\multirow{2}{*}{GaAs} & \multirow{2}{*}{$\text{F}\bar{4}3\textit{m}$} & DFPT & 60 Ha & $6\times6\times6$ & $10\times10\times10$ & GGA-PBE & No \\
			 &  & FD & 408 eV & $\Gamma$ & $4\times4\times4$ & GGA-PBE & No \\
			\multirow{2}{*}{GaP} & \multirow{2}{*}{$\text{F}\bar{4}3\textit{m}$} & DFPT & 60 Ha & $6\times6\times6$ & $10\times10\times10$ & GGA-PBE & Yes \\
			 &  & FD & 408 eV & $\Gamma$ & $4\times4\times4$ & GGA-PBE & No \\
			\multirow{2}{*}{ZnTe} & \multirow{2}{*}{$\text{F}\bar{4}3\textit{m}$} & DFPT & 60 Ha & $6\times6\times6$ & $10\times10\times10$ & GGA-PBE & Yes \\
			 &  & FD & 408 eV & $\Gamma$ & $4\times4\times4$ & GGA-PBE & No \\
			\bottomrule
		\end{tabular}
		\label{tab:params}
	\end{table*}

\begin{table*}[h]
	\centering
	\caption{Structural parameters and valence-electron counts of the four
	chiral crystals. All entries
	are given
	for the right-handed enantiomer, space
	group $\mathrm{P}3_121$ (\#152), whose free positional parameters are
	$3a=(u,0,1/3)$, $3b=(u,0,5/6)$, and $6c=(x,y,z)$. The left-handed structures are their exact mirror images.}
	\begin{tabular}{lcclc}
		\toprule
		Material & $a$ (\AA) & $c$ (\AA) & positional parameters
		& valence electrons \\
		\midrule
		$\alpha$-SiO$_2$ & $5.024102$ & $5.511936$
		& Si $3a$, $u=0.477449$ & Si $4$, O $6$ \\
		& & & O $6c$, $(0.161083,0.745989,0.462881)$ & \\
		Se & $4.516722$ & $5.039358$
		& Se $3a$, $u=0.218929$ & $6$ \\
		Te & $4.502989$ & $5.958131$
		& Te $3a$, $u=0.270844$ & $6$ \\
		$\alpha$-HgS & $4.110968$ & $9.395853$
		& Hg $3a$, $u=0.301067$ & Hg $20$, S $6$ \\
		& & & S $3b$, $u=0.491294$ & \\
		\bottomrule
	\end{tabular}
	\label{tab:struct_psp}
\end{table*}

\FloatBarrier

	The $q$-mesh listed in Table~\ref{tab:params} is the mesh on which the dynamical matrices are computed by DFPT\@. The interatomic force constants obtained from it are Fourier interpolated onto a dense uniform grid on which the Brillouin-zone sums of Eq.~(1) of the main text are evaluated at $T=300$~K\@. That grid is $30\times30\times30$ for $\alpha$-SiO$_2$, Se, and Te, and $36\times36\times36$ for $\alpha$-HgS\@. The $30\times30\times30$ grid is also used for the four achiral crystals.

	At the relaxed lattice constant the calculated band gap of GaAs is $0.15$~eV and its high-frequency dielectric constant reaches $1292$, so the non-analytic term is omitted for GaAs, as in the finite-displacement calculations. Including this term shifts no GaAs phonon frequency by more than $0.8$~cm$^{-1}$ and leaves $G_0$ and $G_u$ at zero.

	The acoustic sum rule and charge neutrality are imposed on the interatomic force constants. The single point $\mathbf{q}=0$, at which the Bose factor of the acoustic modes diverges while $\mathbf{F}_1$ vanishes, is excluded from the sum. Apart from discarding modes with $|\omega|<0.5$~cm$^{-1}$, which are numerically zero, no frequency cutoff is applied, and any
residual negative-frequency mode enters with $|\omega|$. Among the four chiral crystals only $\alpha$-HgS carries any, twenty-four of the $8.4\times10^{5}$ frequencies on its interpolation grid, all on the lowest acoustic branch near $\Gamma$ and all between $-1.0$ and $-2.6$~cm$^{-1}$. Taking $|\omega|$ rather than dropping them changes $G_0$ by $0.22$\% and $G_u$ by $0.16$\%, which leaves the values quoted in the main text unchanged at the precision given.

	$G_0$ and $G_u$ vanish identically by symmetry in the achiral crystals. The computed values are below $2\times10^{-7}$ in magnitude for GaAs, GaP, and ZnTe, and below $10^{-16}$ for Si, where the branch sum vanishes at every wave vector rather than only in the Brillouin-zone sum. Removing the non-analytic correction from the DFPT calculations of GaP and ZnTe changes these values by less than $0.1$\%.

	Convergence with respect to the DFPT $q$-mesh was checked by refining the mesh, requiring both that the dynamical matrices diagonalized directly at the computed $\mathbf{q}$ points give no imaginary frequencies and that $G_0$ be independent of the interpolation grid. The first condition is met by all four chiral crystals with the norm-conserving datasets used here, so it is the $q$-mesh dependence of $G_0$ that selects the meshes listed in Table~\ref{tab:params}, which are the meshes used here. For $\alpha$-HgS that dependence does not settle. The increments do not decelerate over the four meshes tested and the spread of $G_0$ is $0.022$, almost all of it from the innermost shells of $|\mathbf{q}|$, where a handful of points with vanishing $\omega$ carry a Bose weight $\propto1/\omega$ and change sign from mesh to mesh. Outside those shells the two densest meshes agree to better than $0.4$\%.

	\section{Wave-Vector-Resolved Dynamical Chirality}
\begin{figure*}[h]
		\centering
		\makebox[\textwidth][c]{ 
			\includegraphics[width=0.76\textwidth]{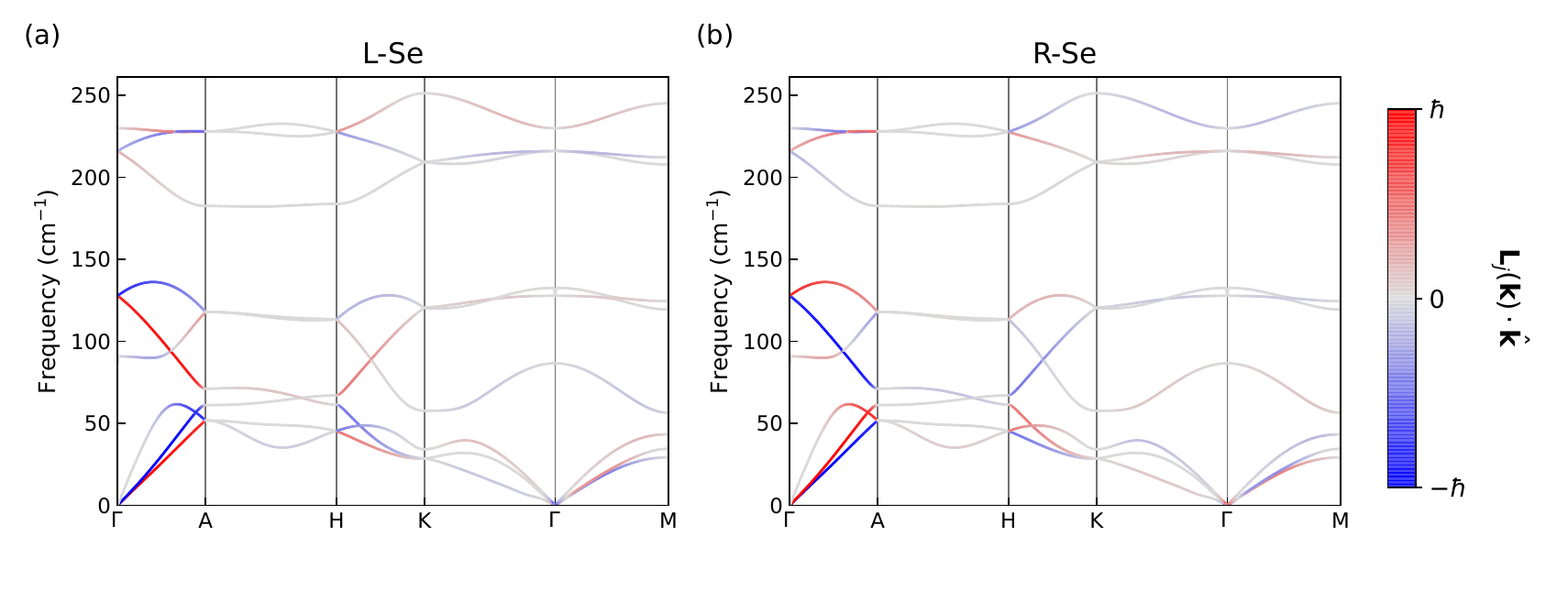}	
		}
		\caption{Phonon dispersions colored by the wave-vector-resolved dynamical chirality $\mathbf{L}_j(\mathbf{k})\cdot\hat{\mathbf{k}}$ in (a) L-Se and (b) R-Se along the paths between high symmetry points. The signs reverse between the enantiomers along corresponding paths.}
		\label{fig:kl_se}
	\end{figure*}

\begin{figure*}[h]
		\centering
		\makebox[\textwidth][c]{ 
			\includegraphics[width=0.76\textwidth]{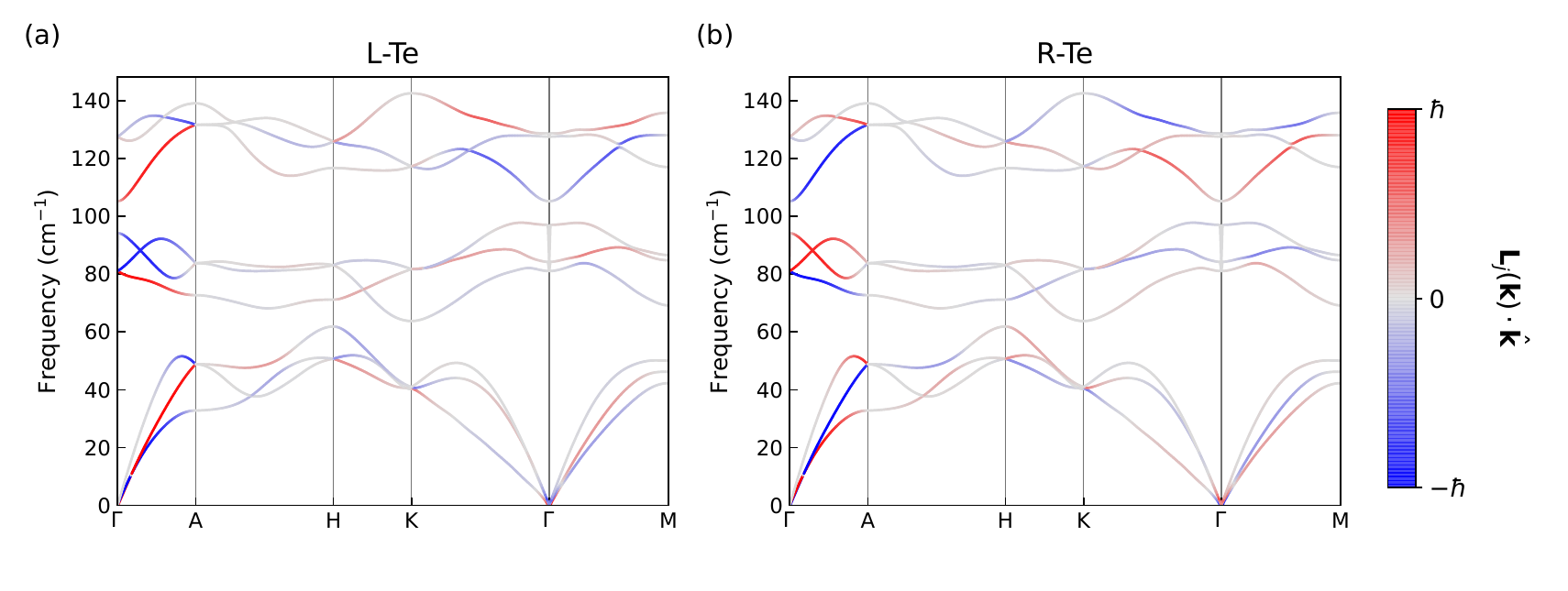}	
		}
		\caption{Phonon dispersions colored by the wave-vector-resolved dynamical chirality $\mathbf{L}_j(\mathbf{k})\cdot\hat{\mathbf{k}}$ in (a) L-Te and (b) R-Te along the paths between high symmetry points. The signs reverse between the enantiomers along corresponding paths.}
		\label{fig:kl_te}
	\end{figure*}

\begin{figure*}[h]
		\centering
		\makebox[\textwidth][c]{ 
			\includegraphics[width=0.76\textwidth]{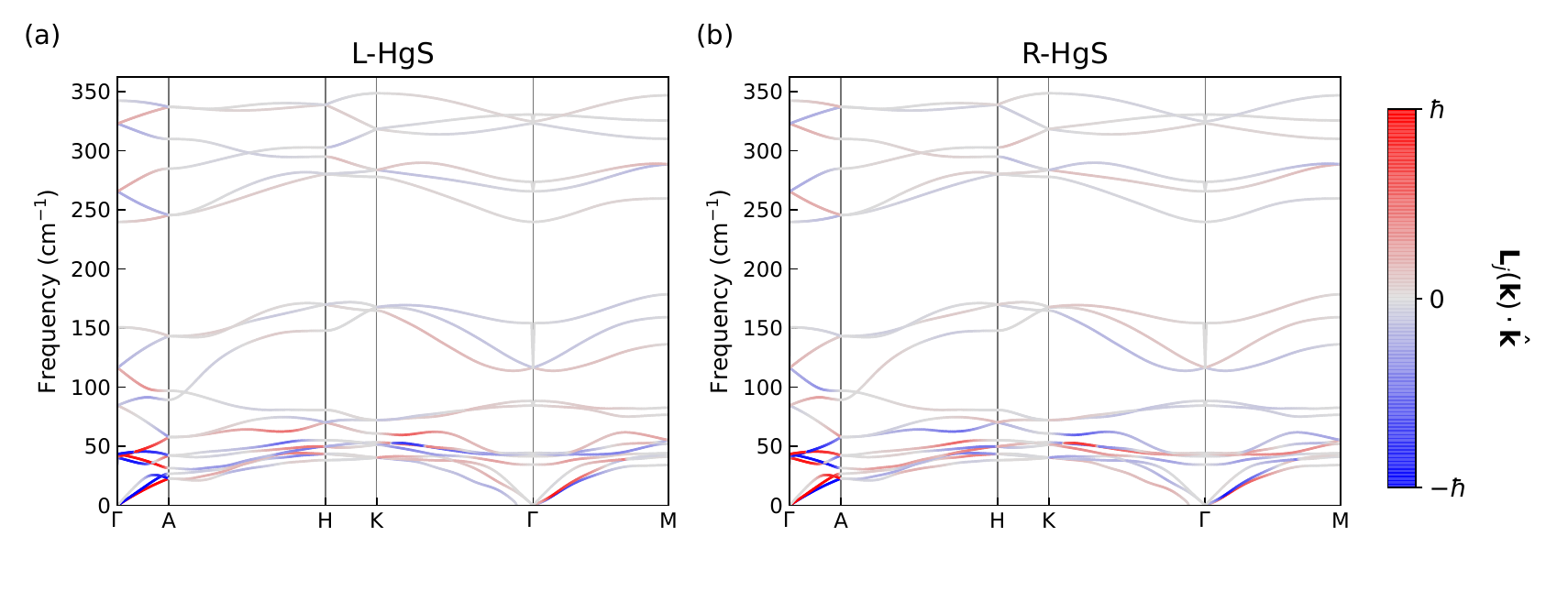}	
		}
		\caption{Phonon dispersions colored by the wave-vector-resolved dynamical chirality $\mathbf{L}_j(\mathbf{k})\cdot\hat{\mathbf{k}}$ in (a) L-HgS and (b) R-HgS along the paths between high symmetry points. The signs reverse between the enantiomers along corresponding paths.}
		\label{fig:kl_hgs}
	\end{figure*}

\begin{figure*}[h]
		\centering
		\makebox[\textwidth][c]{ 
			\includegraphics[width=0.5\textwidth]{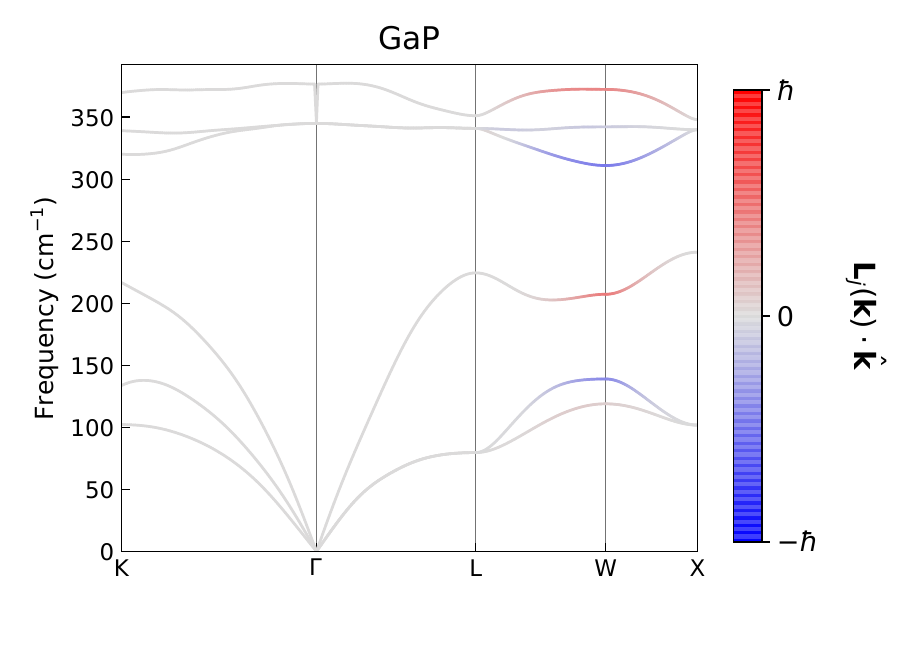}	
		}
		\caption{Phonon dispersions colored by the wave-vector-resolved dynamical chirality $\mathbf{L}_j(\mathbf{k})\cdot\hat{\mathbf{k}}$ in noncentrosymmetric achiral GaP along the paths between high symmetry points. Finite values occur near the W point, although the crystal is not structurally chiral.}
		\label{fig:kl_gap}
	\end{figure*}

\begin{figure*}[h]
		\centering
		\makebox[\textwidth][c]{ 
			\includegraphics[width=0.5\textwidth]{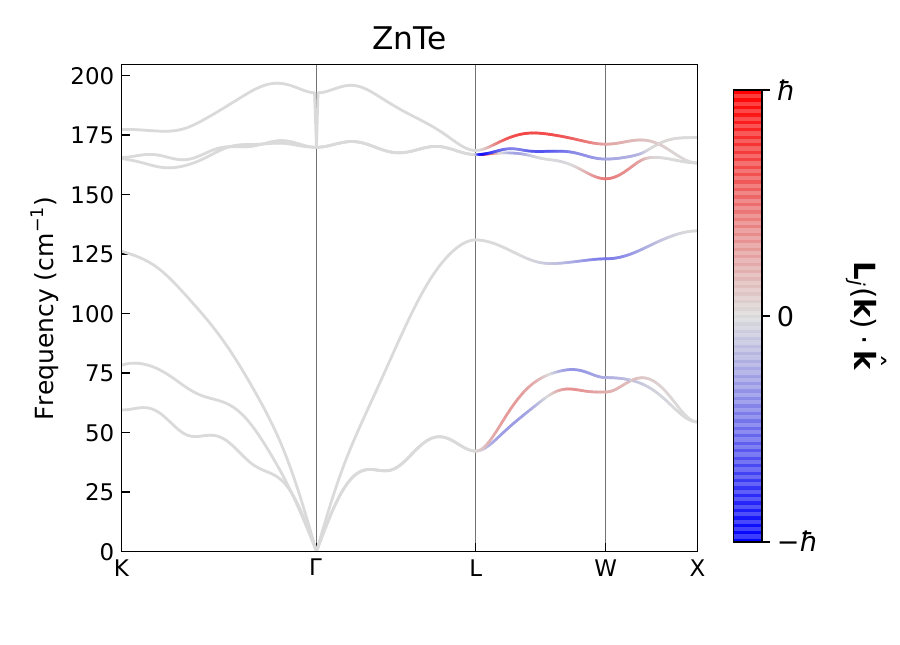}	
		}
		\caption{Phonon dispersions colored by the wave-vector-resolved dynamical chirality $\mathbf{L}_j(\mathbf{k})\cdot\hat{\mathbf{k}}$ in noncentrosymmetric achiral ZnTe along the paths between high symmetry points. Finite values occur near the W point, although the crystal is not structurally chiral.}
		\label{fig:kl_znte}
	\end{figure*}

	The wave-vector-resolved dynamical chirality $\mathbf{L}_j(\mathbf{k})\cdot \hat{\mathbf{k}}$ is computed for all eight materials along high-symmetry paths. In the four chiral crystals it has opposite signs in the left- and right-handed structures, as shown for $\alpha$-SiO$_2$ in Figs.~2(a) and 2(b) of the main text and for Se, Te, and $\alpha$-HgS in Figs.~\ref{fig:kl_se}--\ref{fig:kl_hgs}. In the centrosymmetric achiral crystal Si it vanishes throughout the band structure, as shown in Fig.~2(c) of the main text. In noncentrosymmetric achiral systems, such as GaAs, GaP, and ZnTe, nonzero wave-vector-resolved dynamical chirality can occur despite the absence of structural chirality. The result for GaAs is shown in Fig.~2(d) of the main text, while Figs.~\ref{fig:kl_gap} and \ref{fig:kl_znte} show those for GaP and ZnTe, respectively.

	\newpage

	\textbf{Chiral Materials}
%
%
	\clearpage
	
	\textbf{Achiral Materials}
	
	\clearpage

    \section{Temperature dependence of bulk dynamical chirality as characterized by $G_0$ and $G_u$}
    As shown in Fig.~\ref{fig:linear_T_G}, the magnitudes of $G_0$ and $G_u$ grow approximately linearly with temperature for the chiral materials shown. In the harmonic approximation the frequencies and the eigenvectors are independent of temperature, so the whole temperature dependence of Eq.~(1) of the main text enters through the occupation factor
\begin{equation}
  f(\omega)=\frac{1}{e^{\hbar\omega/k_BT}-1}+\frac{1}{2}
  =\frac{1}{2}\coth\!\left(\frac{\hbar\omega}{2k_BT}\right)
  \simeq\frac{k_BT}{\hbar\omega} ,
\end{equation}
in which the $1/2$ term is the zero-point contribution to the phonon angular momentum~\cite{Zhang2014}. Summed over the branches at a given $\mathbf{k}$ that term cancels, since $\sum_j \mathbf{L}_j(\mathbf{k})=0$. That cancellation is exact for the complete branch sum. The sums evaluated here discard the modes of numerically vanishing frequency described in Sec.~I, and they still give zero at $T=0$. The next term of the high-temperature expansion is $\hbar\omega/(12k_BT)$, so no constant survives, and replacing $f$ by $k_BT/\hbar\omega$ in Eq.~(1) of the main text makes both $G_0$ and $G_u$ proportional to $T$. That step requires $k_BT\gg\hbar\omega_j(\mathbf{k})$ for the modes that dominate the sums. At $300$~K this holds for the low-frequency branches, which carry most of the weight through the Bose factor, but not for the highest ones, so the linearity is approximate.
    
\clearpage
\refstepcounter{figure}
\label{fig:linear_T_G}
\begin{center}
	\includegraphics[width=0.75\textwidth]{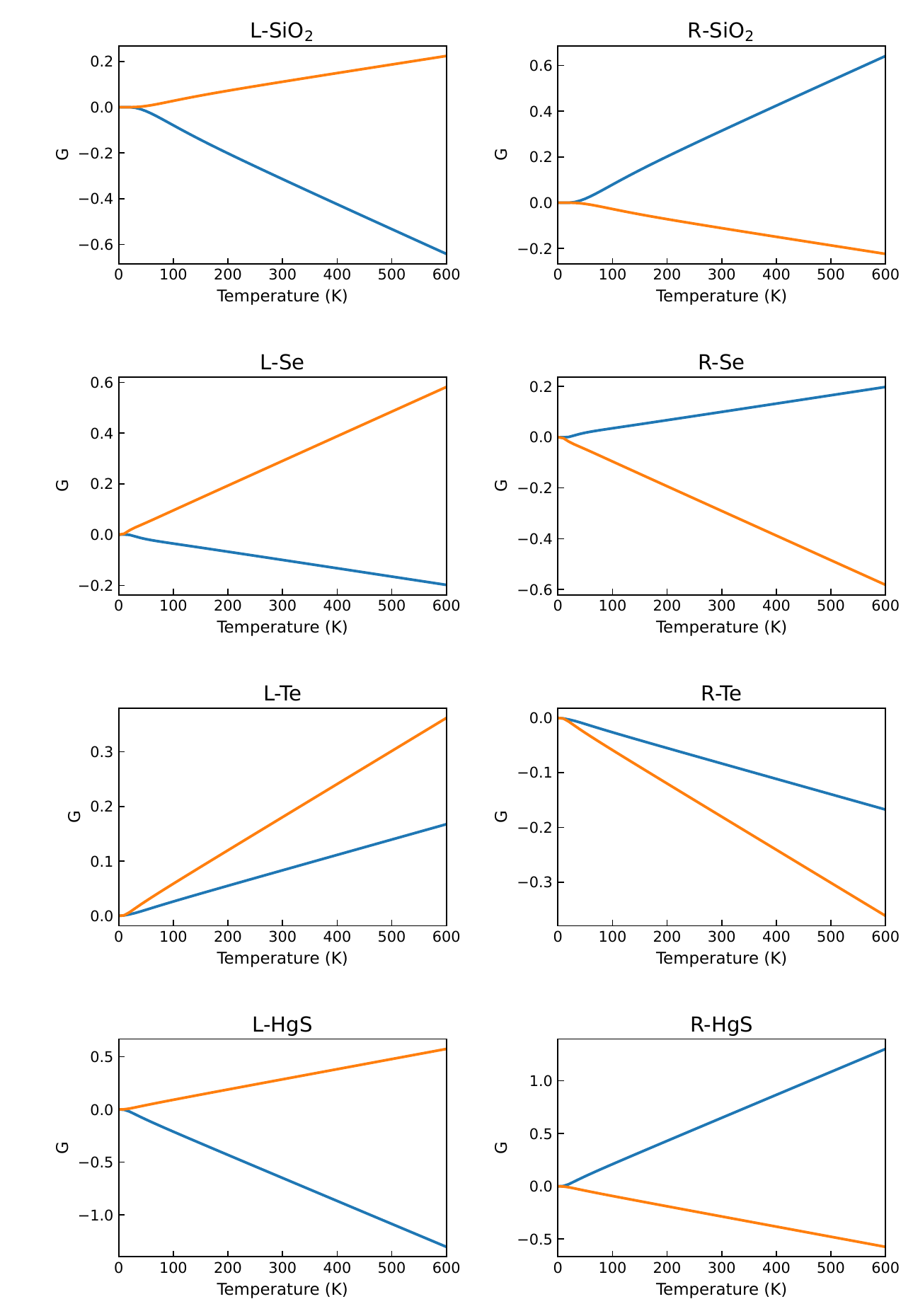}

	\textbf{FIG. \thefigure.} Temperature dependence of the bulk dynamical chirality measures $G_0$ (blue) and $G_u$ (orange). Each row is one material, with the left-handed enantiomer in the left column and the right-handed one in the right column. Both measures reverse sign between the enantiomers and vary approximately linearly with temperature over the range shown, and both vanish at $T=0$ within numerical accuracy.
\end{center}
    
	\newpage

	\section{Derivation of Structure Factor $F_1(\mathbf{k})$}

	We first give the lowest-order periodic structure factors for $D_3$, $O_h$, and $T_d$~\cite{IshitoTe, Hayami2024}, then derive the $D_3$ expression by symmetry projection and state the normalization convention.
	
	\noindent For $D_3$, the components of $\mathbf{F}_1(\mathbf{k})$ are given by
	{\footnotesize
    \begin{equation}
    \begin{aligned}
    F_1^x(\mathbf{k}) &= \frac{2}{\sqrt{3}}
    \left[ 2 \cos \left(\frac{k_x a}{2}\right)
    + \cos \left(\frac{\sqrt{3} k_y a}{2}\right) \right]
    \sin \left(\frac{k_x a}{2}\right), \\[0.8em]
    F_1^y(\mathbf{k}) &= 2 \cos \left(\frac{k_x a}{2}\right)
    \sin \left(\frac{\sqrt{3} k_y a}{2}\right), \\[0em]
    F_1^z(\mathbf{k}) &= \sin \left(k_z c\right) .
    \end{aligned}
    \label{eq:f1d3}
    \end{equation}
    }
	while for $O_h$ and $T_d$, they take the form
	{\footnotesize
    \begin{align}
    \mathbf{F}_1 (\mathbf{k}) = \left(\sin(k_x a),\sin(k_y a),\sin(k_z a)\right).
    \label{eq:f1cubic}
    \end{align}
    }
	
	\noindent To derive the $D_3$ structure factor, we use symmetry projection to identify the allowed components of a periodic Fourier basis. We first expand the $\mathbf{k}$-dependent angular momentum $\mathbf{L}_j(\mathbf{k})$ as
	\begin{align}
		\mathbf{L}_j(\mathbf{k})=c_1(j)\mathbf{F}_1(\mathbf{k})+c_2(j)\mathbf{F}_2(\mathbf{k})+...,
	\end{align}
	where $\mathbf{F}_i(\mathbf{k})$ represent basis functions with different periodicities and $\mathbf{F}_1(\mathbf{k})$ corresponds to the lowest-order term. Next, applying the projection operator from group theory to $\mathbf{L}_j(\mathbf{k})$ returns its component in the chosen irreducible representation,
	\begin{align}
		\mathcal{P}_{\alpha\beta}[\mathbf{L}_j(\mathbf{k})] = \sum_i c_{i\beta}(j) F_i^{\alpha}(\mathbf{k}),
	\end{align}
	where $\alpha, \beta = x, y, z$ denote Cartesian components, $c_{i\beta}(j)$ is the $\beta$ component of $\mathbf{L}_j$ on the $i$th basis set, and the sum runs over those basis functions of the expansion above that belong to that representation. The projection does not by itself single out one of them. Here $\mathbf{F}_1$ denotes the lowest-order member, and it is the function retained throughout.

	The projection operator is defined by
	\begin{align}
		\mathcal{P}_{\alpha\beta} \equiv \frac{d}{g} \sum_{\mathcal{G}} D^*_{\alpha\beta}(\mathcal{G}) \mathcal{G}, 
	\end{align}
	where $d$ is the dimension of irreducible representation, $g$ is the number of symmetry operations, $\mathcal{G}$ is the symmetry operation, and $D_{\alpha\beta}(\mathcal{G})$ is the representation matrix of $\mathcal{G}$.
	
\begin{figure}[h]
		\makebox[\textwidth][c]{ 
			\includegraphics[width=0.6\textwidth]{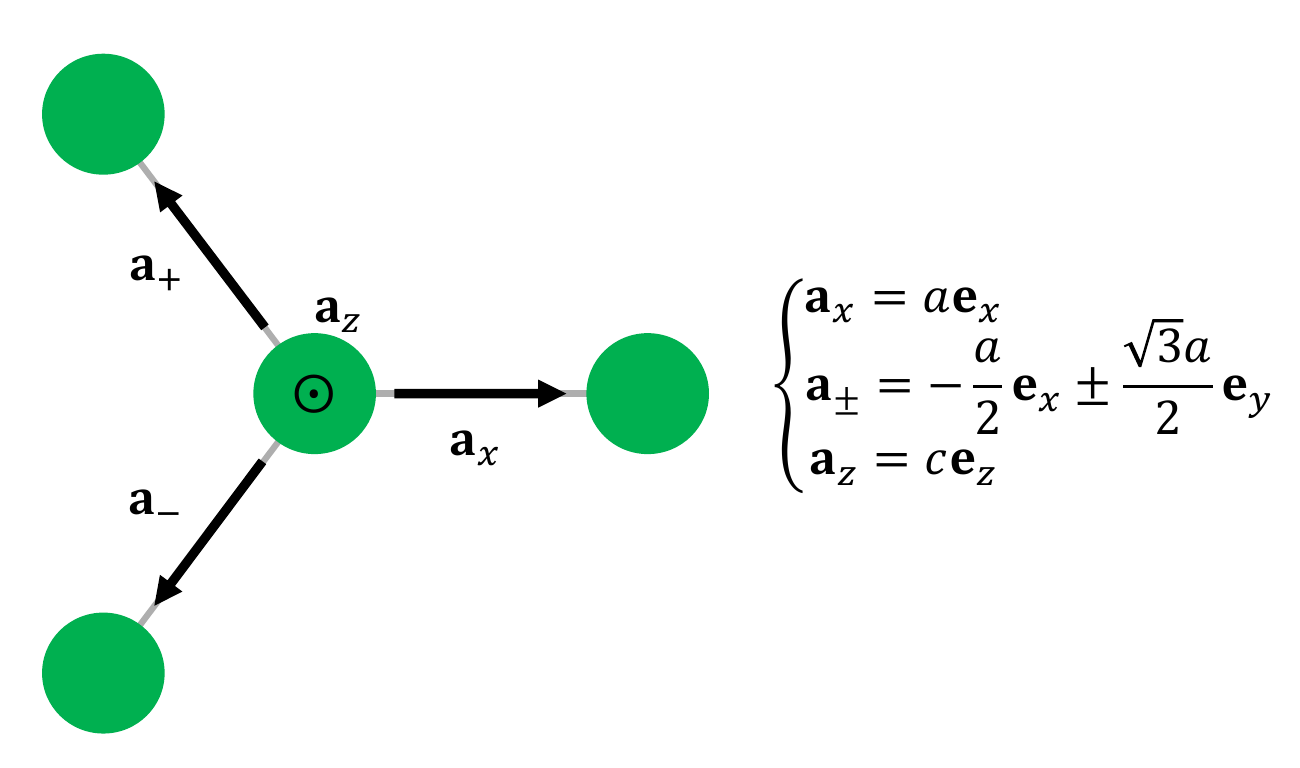} 
		}
		\caption{Schematic of the basal-plane vectors $\mathbf{a}_x$, $\mathbf{a}_+$, and $\mathbf{a}_-$ and the $c$-axis vector $\mathbf{a}_z$ of the $D_3$ materials, with $\mathbf{a}_x$ along $x$. These vectors define the Cartesian frame used for the phase factors and for $\mathbf{F}_1(\mathbf{k})$.}
		\label{fig:unit_cell}
	\end{figure}

	\noindent The $D_3$ character table (Table~\ref{tab:irrep_cell}) and the vectors in Fig.~\ref{fig:unit_cell} specify the transformations used in the projection.
	\begin{table}[h]
		\centering
		\renewcommand{\arraystretch}{1.5}
		\begin{tabular}{c c c c c c c}
			\toprule
			irrep. & $1$ & $C_2$ & $C_2$ & $C_2$ & $C_3$ & $C_3^{-1}$ \\
			\midrule
			$A_1$ & 1 & 1 & 1 & 1 & 1 & 1 \\
			$A_2$ & 1 & $-1$ & $-1$ & $-1$ & 1 & 1 \\
			\midrule
			$E$ & 
			$\begin{pmatrix} 1 & 0 \\ 0 & 1 \end{pmatrix}$ & 
			$\begin{pmatrix} 1 & 0 \\ 0 & -1 \end{pmatrix}$ & 
			$\begin{pmatrix} -\frac{1}{2} & -\frac{\sqrt{3}}{2} \\ -\frac{\sqrt{3}}{2} & \frac{1}{2} \end{pmatrix}$ & 
			$\begin{pmatrix} -\frac{1}{2} & \frac{\sqrt{3}}{2} \\ \frac{\sqrt{3}}{2} & \frac{1}{2} \end{pmatrix}$ & 
			$\begin{pmatrix} -\frac{1}{2} & -\frac{\sqrt{3}}{2} \\ \frac{\sqrt{3}}{2} & -\frac{1}{2} \end{pmatrix}$ &
			$\begin{pmatrix} -\frac{1}{2} & \frac{\sqrt{3}}{2} \\ -\frac{\sqrt{3}}{2} & -\frac{1}{2} \end{pmatrix}$ \\
			\midrule
			$\chi_E$ & 2 & 0 & 0 & 0 & $-1$ & $-1$ \\
			\bottomrule
			\toprule
			$\mathcal{G}[\mathbf{a}_x]$ & $\mathbf{a}_x$ & $\mathbf{a}_x$ & $\mathbf{a}_-$ & $\mathbf{a}_+$ & $\mathbf{a}_+$ & $\mathbf{a}_-$ \\
			$\mathcal{G}[\mathbf{a}_z]$ & $\mathbf{a}_z$ & $-\mathbf{a}_z$ & $-\mathbf{a}_z$ & $-\mathbf{a}_z$ & $\mathbf{a}_z$ & $\mathbf{a}_z$ \\
			$\mathcal{G}[p_x]$ & $p_x$ & $p_x$ & $p_-$ & $p_+$ & $p_+$ & $p_-$ \\
			$\mathcal{G}[p_z]$ & $p_z$ & $p_z^*$ & $p_z^*$ & $p_z^*$ & $p_z$ & $p_z$ \\
			\bottomrule
		\end{tabular}
		\caption{Irreducible representations of $D_3$, the symmetry operations acting on the unit cell vectors, and the same operations acting on the phase factors of those vectors.}
		\label{tab:irrep_cell}
	\end{table}

	Applying the projection operator to the unit cell vectors of Fig.~\ref{fig:unit_cell} gives zero for $A_1$, the $z$ direction for $A_2$, and the $x$ and $y$ directions for the two partners of $E$. In $D_3$ a polar vector therefore decomposes as $A_2$ for its $c$-axis component and $E$ for its basal-plane pair.

    We next apply the same projection procedure to the angular momentum with periodicity in the Brillouin zone. To construct the periodic basis functions we introduce the phase factors $p_x = e^{i\mathbf{k} \cdot \mathbf{a}_x}$, $p_{\pm} = e^{i\mathbf{k} \cdot \mathbf{a}_{\pm}}$, and $p_z = e^{i\mathbf{k} \cdot \mathbf{a}_z}$, together with their symmetry-related counterparts. Their transformation under the symmetry operations is listed in Table~\ref{tab:irrep_cell}. The axial combination $p_z-p_z^*=2i\sin k_zc$ transforms as $A_2$, while the two independent basal-plane combinations $2p_x-(p_++p_-)$ and $p_+-p_-$ transform as $E$. The $A_1$ sector does not contribute to the vector basis, which transforms as $A_2\oplus E$ under $D_3$.

	Taking the parts odd in $\mathbf{k}$ and applying the normalization convention specified below gives $F_1^z$ from $A_2$ and $(F_1^x,F_1^y)$ from $E$, which are the components written in Eq.~(\ref{eq:f1d3}). Explicitly,
	\[
	F_1^x(\mathbf{k})=\frac{1}{\sqrt{3}}\,\mathrm{Im}\,(2p_x-p_+-p_-),\qquad
	F_1^y(\mathbf{k})=\mathrm{Im}\,(p_+-p_-),\qquad
	F_1^z(\mathbf{k})=\mathrm{Im}\,p_z ,
	\]
	which give Eq.~(\ref{eq:f1d3}) for $\mathbf{a}_x=a(1,0,0)$, $\mathbf{a}_\pm=a(-1/2,\pm\sqrt{3}/2,0)$, and $\mathbf{a}_z=c(0,0,1)$. The corresponding polar-vector periodic bases for $O_h$ and $T_d$ can be constructed by symmetry and are given in Eq.~(\ref{eq:f1cubic}).

	The coefficients of the in-plane components are fixed by normalization over the entire Brillouin zone, $\langle (F_1^x)^2\rangle=\langle (F_1^y)^2\rangle=1$, while the axial component is taken as $F_1^z=\sin(k_z c)$, for which $\langle (F_1^z)^2\rangle=1/2$. The same convention is used for all materials, so that the relative weight of the axial and in-plane channels entering $G_0$ and $G_u$ is fixed throughout. In the cubic crystals the three components of Eq.~(\ref{eq:f1cubic}) are equivalent by symmetry and each has mean square $1/2$. Symmetry fixes which functions are allowed but not their relative normalization, so this is a convention of the present work rather than a property of the crystal. The absolute values of $G_0$ and $G_u$ depend on it, and so does the sign of $G_0$ wherever the axial and basal-plane contributions nearly cancel, as they do for R-Se.

	In a cubic crystal the three diagonal components are equal by symmetry, so $G_u$ vanishes identically whether or not the crystal is chiral. A vanishing $G_u$ therefore does not indicate the absence of chirality, and it is $G_0$ that carries that information.

	In practice, $\mathbf{L}_j(\mathbf{k})$ and $\mathbf{F}_1(\mathbf{k})$ must be expressed in the same Cartesian frame, that of Fig.~\ref{fig:unit_cell}, in which the first lattice vector lies along $x$. When the input cell follows a different setting (as for the conventional hexagonal cell of $\alpha$-quartz, whose first lattice vector makes an angle of $-30^\circ$ with $x$), the phonon angular momentum is rotated into the $\mathbf{F}_1$ frame before the inner product is evaluated. The axial combination $(G_0+2G_u)/3$ reduces to the sum of
$L^z_j F^z_1$ alone, so it is invariant under rotations about $c$ and tests only
the axial part, not the alignment in the plane.
	
\section{Alternative validation and computational details}
\label{sec:vasp}

To cross-validate the proposed measures of phonon chirality, we performed calculations using an alternative first-principles approach based on the finite-displacement (FD) method, as implemented in the \textsc{phonopy} package~\cite{Togo2015} in combination with the \textsc{vasp} code~\cite{Kresse1996}.

\subsection{Computational setup}

The FD calculations use the projector augmented-wave method, with the exchange-correlation functional of each crystal as listed in Table~\ref{tab:params}. The datasets are the standard ones distributed with \textsc{vasp}, identified by their \texttt{TITEL} lines as Si 05Jan2001 and O 08Apr2002 for $\alpha$-SiO$_2$, Se 06Sep2000 for Se, Te 08Apr2002 for Te, Hg 06Sep2000 and S 06Sep2000 for $\alpha$-HgS, Si 05Jan2001 for Si, Ga 08Apr2002 and As 22Sep2009 for GaAs, Ga 08Apr2002 and P 06Sep2000 for GaP, and Zn 06Sep2000 and Te 08Apr2002 for ZnTe. The aspherical contributions inside the augmentation spheres were included for Se and left at their default for the other materials, and the effect of that difference has not been assessed. The FD calculations were performed on the same structural cells as the DFPT calculations for all eight materials. The structures are described in Sec.~\ref{sec:comp_params}, with the chiral structural parameters listed in Table~\ref{tab:struct_psp}. Atomic displacements of $0.01~\text{\AA}$ were employed in the finite-displacement calculations.

Phonon properties were computed using \textsc{phonopy}, and $G_0$ and $G_u$ were evaluated from the resulting eigenfrequencies and eigendisplacements by summing over a uniform $19\times19\times19$ $q$-point mesh at $T=300~\mathrm{K}$. For the chiral crystals the FD and the DFPT calculations use the same Bose occupation factor $f=n_\text{B}+1/2$.

	The non-analytic (LO--TO) term was included for the chiral crystals through their Born effective charges and dielectric tensors and omitted for the four achiral crystals, where $G_0$ and $G_u$ vanish identically by symmetry. For $\alpha$-SiO$_2$, Se, and Te this term was evaluated with the default scheme of \textsc{phonopy}, and for $\alpha$-HgS with its alternative scheme, the default producing many negative-frequency modes for that crystal. For R-SiO$_2$, R-Se, and R-Te the FD calculations use the Born effective charges and dielectric tensors of the DFPT calculations, whereas for R-HgS these tensors are computed separately with the same functional. The Born effective charges taken from the DFPT calculations are transposed on input to \textsc{phonopy}, whose index order is the transpose of the \textsc{anaddb} one.

	The Brillouin-zone sum excludes $\mathbf{q}=0$ and any residual negative-frequency mode, and no frequency cutoff is applied except for the four achiral crystals, which retain a $0.3$~THz cutoff. There the cutoff is immaterial, since $G_0$ and $G_u$ vanish identically by symmetry. Dropping the negative modes differs from Sec.~I, where they enter with their absolute value. Se, Te, and R-HgS carry no negative-frequency modes, nor does R-SiO$_2$ at the larger supercell. At the smaller R-SiO$_2$ supercell twelve such modes occur, and the two conventions differ by $0.05$ in $G_0$.

\subsection{Consistency with DFPT results}

The calculations using the alternative approach reproduce the key qualitative features reported in the main text. The wave-vector-resolved dynamical chirality changes sign between enantiomers across the Brillouin zone, centrosymmetric crystals exhibit vanishing chirality at all wave vectors, and noncentrosymmetric achiral crystals show finite local chirality that cancels in the Brillouin-zone sum, resulting in vanishing bulk dynamical chirality~\cite{Yang2026}.

Near $\Gamma$ along $\Gamma$--A the two transverse-acoustic branches of both Se and Te carry opposite angular momentum approaching $\pm\hbar$. The two methods agree on $|\ell^z|$ to $0.001$ at reduced wave numbers of $0.05$ and $0.10$, although the transverse-acoustic frequencies themselves differ by up to $1.4$~cm$^{-1}$ there. The magnitude falls away from $\Gamma$, and at a reduced wave number of $0.4$ it is $0.88$ and $0.89$ on the two branches of Se and $0.4$ on one of the two branches of Te, so the statement is local to $\Gamma$. The branches have to be followed by their polarization rather than by their order in frequency, which differs between the two methods and between the interpolated and the directly computed dynamical matrix.

These results confirm that the symmetry-based characterization of phonon chirality is robust across different computational approaches. For Se and Te the FD approach reproduces the sign of $G_u$ at every supercell, negative for both. It reproduces the sign of $G_0$ at every supercell for Te and, for Se, only once the supercell has more than four cells along $c$, the shorter one giving the opposite sign. At the supercells quoted in Table~\ref{tab:comparison} the finite-displacement values reproduce the sign of both measures for all four chiral crystals.

\subsection{Quantitative differences}
	While the qualitative behavior is consistent, quantitative differences in the values of $G_0$ and $G_u$ are observed between the finite-displacement and density-functional perturbation theory (DFPT) calculations used in the main text. Representative comparisons are summarized in Table~\ref{tab:comparison}. Both calculations of this comparison were set up from the same relaxed cell for each crystal, the two input files agreeing to $0.01$~ppm in the lattice vectors, so the differences do not arise from the structure within that comparison.

	The supercell series are $4\times4\times6$, $5\times5\times4$, and $5\times5\times6$ for Se, $4\times4\times4$ and $5\times5\times4$ for Te, $3\times3\times3$ and $4\times4\times3$ for R-SiO$_2$, and $3\times3\times4$ and $4\times4\times4$ for $\alpha$-HgS. For Se only the supercells with an even number of divisions along $c$ are used, since the A point lies on the supercell grid only for an even division. The uncertainties quoted for $G_0$ and $G_u$ in Table~\ref{tab:comparison} are the spread of the supercell series rather than a numerical precision.

	Both $G_0$ and $G_u$ are small residuals of large, mutually cancelling acoustic and optical contributions, and are therefore sensitive to the real-space range of the interatomic force constants captured by the FD supercell. The supercell also sets the effective electronic $k$-point density, so the two effects are not separated here. $G_u$ is the shallower cancellation, its supercell spread being smaller than that of $G_0$ by factors of two to eight. For Se and Te the DFPT value lies within the span of the supercell series. These differences may reflect the construction of the force constants, the pseudopotentials, and the finite real-space range of the FD supercells, whose separate contributions have not been established.

	For the R-SiO$_2$ entry of Table~\ref{tab:comparison}, pseudopotential choice remains important, as tests on the unrelaxed cell showed. The projector augmented-wave dataset used in the earlier \textsc{abinit} calculation does not converge for this structure, the augmentation spheres of Si and O overlapping and the softest mode varying by more than $120$~cm$^{-1}$ with the compensation-charge cutoff. The value quoted here uses a dataset whose exchange-correlation functional matches the one used in the calculation, and replacing the earlier dataset by it raises $G_0$ by $60$\%.
	$G_0$ of R-SiO$_2$ is dominated by the acoustic branches, whose Bose weight $\propto 1/\omega$ makes it sensitive to any systematic softening of the low-frequency spectrum, and that softening differs strongly between pseudopotential sets, so that the spread between methods reported here is a lower bound on the sensitivity rather than a converged comparison. The non-analytic term is not the amplitude-limiting factor, the supercell controlling the convergence while the sign is robust against the NAC treatment, the integration mesh, and the occupation convention.

	Across the four chiral crystals for which both methods are available the FD and DFPT values differ by at most $0.05$ in $G_0$, for R-Se, and by at most $0.08$ in $G_u$, for R-HgS\@. The smaller tabulated difference for R-HgS, from $0.12$ to $0.04$ in $G_0$, follows from replacing the single-supercell value by the centre of two supercells, and it establishes neither better agreement between the methods nor improved convergence. For R-HgS both calculations use GGA-PBE, so the functional is not the source. Two FD supercells, $3\times3\times4$ and $4\times4\times4$, have been evaluated for that crystal, both with four cells along $c$, so the convergence along that axis has not been established, and the PBEsol-relaxed cell on which both calculations are run is not stationary under PBE\@. Neither effect has been isolated as the cause.

	On the undisplaced $4\times4\times4$ R-HgS supercell the residual force under PBE is $0.13$~eV/\AA, against $0.009$~eV/\AA\ under PBEsol. Subtracting that offset changes $G_0$ and $G_u$ by less than $10^{-5}$, and replacing the Born effective charges and dielectric tensor by those obtained with the other functional moves them by less than $0.001$.

	Overall, the agreement in sign and magnitude documented above supports the proposed framework, while the spread within each supercell series indicates that absolute values of $G_0$ and $G_u$ should be compared only within a consistent computational setup.

\begin{table}[htbp]
\caption{Comparison of the bulk dynamical chirality measures $G_0$ and $G_u$ obtained from DFPT and finite-displacement (FD) calculations. DFPT values are the main-text results of Table~I\@.}
\begin{tabular}{lcccccc}
\toprule
Material & $G_0$ (DFPT) & $G_0$ (FD) & $|\Delta G_0|$ & $G_u$ (DFPT) & $G_u$ (FD) & $|\Delta G_u|$ \\
\midrule
R-SiO$_2$ & $+0.31$ & $+0.30\pm0.07$ & $0.02$ & $-0.11$ & $-0.10\pm0.02$ & $0.01$ \\
R-Se & $+0.10$ & $+0.05\pm0.06$ & $0.05$ & $-0.29$ & $-0.30\pm0.03$ & $0.01$ \\
R-Te & $-0.08$ & $-0.07\pm0.05$ & $0.01$ & $-0.18$ & $-0.17\pm0.01$ & $0.01$ \\
R-HgS & $+0.65$ & $+0.61\pm0.08$ & $0.04$ & $-0.29$ & $-0.20\pm0.01$ & $0.08$ \\
Si        & 0.00 & 0.00 & & 0.00 & 0.00 & \\
GaAs      & 0.00 & 0.00 & & 0.00 & 0.00 & \\
GaP      & 0.00 & 0.00 & & 0.00 & 0.00 & \\
ZnTe     & 0.00 & 0.00 & & 0.00 & 0.00 & \\
\bottomrule
\end{tabular}
\vspace{2pt}
\begin{flushleft}
\footnotesize
$|\Delta G_0|$ and $|\Delta G_u|$ come from the unrounded values and need not equal the difference of the rounded entries. Where a supercell series is available, the FD entry is its centre $(\max+\min)/2$ and the spread its half-width $(\max-\min)/2$, which is neither a statistical error nor a bound on the convergence error.
\end{flushleft}
\label{tab:comparison}
\end{table}

\FloatBarrier
\section{Linear-in-$k$ splitting of the chiral doublets}
\label{sec:split}

\begingroup
\subsection{Symmetry of the split optical $E$ doublet}

To describe how an optical $E$ doublet splits along $\mathbf{k}\parallel c$
we need a label that tells its two branches apart at every $k_z$. The
threefold screw provides one.

For $\mathbf{k}\parallel c$ the little co-group, the point operations that
leave the wave vector unchanged, is $C_3$. It is generated by the rotation
part of the threefold screw $g=\{C_{3z}\mid\mathbf{c}/3\}$ of $P3_121$, the
right-handed enantiomer studied here, that is, a $120^\circ$ rotation about
$c$ followed by a translation of $\mathbf{c}/3$. The left-handed enantiomer
has $\{C_{3z}\mid2\mathbf{c}/3\}$ in $P3_221$. We write $S$ for the
representation of $g$ on the mass-weighted displacements at $\Gamma$, the
matrix by which the screw acts on them.

At $\Gamma$ the two partners of the doublet pick up different phases under
$S$, $S|\pm\rangle=e^{\mp2\pi i/3}|\pm\rangle$, and this is the label: the
crystal angular momentum $m=\pm1$. We write
$\lambda_\pm(k_z)=\omega_\pm^2(k_z)$ for the two exact dynamical-matrix
eigenvalues that connect continuously to that doublet.

Away from $\Gamma$ the screw acts through the wave-vector-dependent
representation $U_{k_z}(g)$, whose eigenvalues pick up a further Bloch
phase from the translation $\mathbf{c}/3$. That phase is the same for both
partners, so the ratio of their eigenvalues is unchanged and the two stay
distinct. The label $m$ therefore holds along the whole line, and because
the Bloch phase enters only the screw eigenvalues, the frequencies and the
$\lambda_\pm$ stay real.

Time reversal relates the two branches as $\lambda_+(k_z)=\lambda_-(-k_z)$. Hence
$a(k_z)=[\lambda_+(k_z)+\lambda_-(k_z)]/2$ is even and $b(k_z)=[\lambda_+(k_z)-\lambda_-(k_z)]/2$
is odd in $k_z$, both being real~\cite{PineDresselhaus1969}. Because the two screw sectors do not
mix, as Eq.~(\ref{eq:commute}) below states, the two-branch spectrum is
represented by the effective diagonal matrix
\begin{equation}
D_{\mathrm{eff}}(k_z)=a(k_z)\tau_0+b(k_z)\tau_z ,
\end{equation}
which is not the same object as the fixed projection $P_ED_\mathrm{full}(k_z)P_E$.
Expanding $a=\omega_0^2+a_2k_z^2+O(k_z^4)$ and $b=b_1k_z+O(k_z^3)$ and taking the
square root of $\lambda_\pm=a\pm b$ gives Eq.~(2) of the main text, with
$B=b_1/(2\omega_0)$, where $a_2$ is the coefficient of
the exact half-sum. To the same order the branches are $\omega_\pm=\omega_0\pm Bk_z+Ak_z^2+O(k_z^3)$ with $A=(a_2-B^2)/(2\omega_0)$. The quadratic term is common to the two branches, which is why the difference in Eq.~(2) of the main text carries no term in $k_z^2$.

What forbids mixing between the two screw sectors is the symmetry of
the dynamical matrix itself. Along the axis
\begin{equation}
[D(k_z),U_{k_z}(g)]=0 ,
\label{eq:commute}
\end{equation}
so the two-dimensional subspace that the doublet spans is invariant under
$U_{k_z}(g)$, and its two branches carry distinct screw eigenvalues. We have
verified both statements numerically for all four chiral crystals.

The coefficients $a_2$ and $b_1$ above are therefore read from the exact
eigenvalues $\lambda_\pm$, and not from the fixed $\Gamma$-point projection
$P_ED_\mathrm{full}(k_z)P_E$, which is a different object. The two differ
because $U_{k_z}(g)$ and $U_\Gamma(g)$ are related by a diagonal Bloch
phase that is not proportional to the identity, so they need not share an
eigenbasis. The fixed projection then need not be diagonal in the basis
used here, and coupling to other modes of the same screw sector enters at
order $k_z^2$, so its quadratic coefficient need not be the exact $a_2$.

	These expansions require $D(k_z)$ to be analytic at $k_z=0$, which the
non-analytic dipole--dipole term could spoil. Along this line that term is
$Z^*_{z\alpha}Z^*_{z\beta}/\varepsilon_{zz}$: the factors of $|\mathbf{k}|$
in its numerator and denominator cancel, so it is independent of $k_z$ and
adds only a constant, leaving the expansion analytic. Its projection on an
optical $E$ doublet vanishes in any case, because $z$, and with it the mode
effective charge along $c$, transforms as $A_2$ rather than $E$. Nor does
the off-diagonal form of the atomic $Z^*$ by itself produce a coupling to
the $E$ doublets: for Se, Te, and $\alpha$-SiO$_2$ the summed
$|Z^*_{\nu z}|^2$ of every optical $E$ doublet is smaller than that of the
$A$ singlets by at least fifteen orders of magnitude.

	The transverse-acoustic doublet is the one case in
which $b_1$ is forced to vanish. There $a(0)=0$, so $\lambda_\pm=a_2k_z^2\pm b_1k_z$ to
leading order, and a nonzero $b_1$ would make $\lambda_-$ negative at small
$|k_z|$ of one sign. Hence the two sound velocities coincide and the
splitting of that doublet starts at $O(k_z^2)$.

In the achiral group $C_{3v}$ the vertical mirrors also belong to the little
co-group. A mirror exchanges $|+\rangle$ and $|-\rangle$, so it anticommutes
with $\tau_z$ while commuting with $\tau_0$. Requiring $D$ to commute with the
mirror therefore sets $b\equiv0$, and the doublet does not split at any order.
\endgroup

\subsection{Extraction of the splitting slopes}

The splitting slope $B$ of Eq.~(2) of the main text is obtained from the
same dynamical matrices as $G_0$ and $G_u$. The matrices are interpolated along
$\mathbf{q}=(0,0,\zeta)$ and $\omega_+-\omega_-$ is fitted over
$\zeta\le0.04$, with the doublet followed through its projection onto the
degenerate subspace at $\Gamma$. Branch indices cannot be used for this,
because along $\mathbf{k}\parallel c$ the longitudinal $A_2$ modes move through
the ordering of the frequencies. For $\alpha$-SiO$_2$ the backscattering wave number of Table~\ref{tab:raman} corresponds to $\zeta\simeq0.003$, well inside the fitted range $\zeta\le0.04$. For the lowest Te doublet $E^{(1)}$ the slope is constant to four digits over the wave numbers reached in backscattering Raman, which is stability against the fitting window and not convergence with respect to the $q$-mesh. The
splitting of that doublet is largest near $\zeta\approx0.15$ and collapses
beyond it, where it anticrosses other branches.

Table~\ref{tab:doublets} lists every optical $E$ doublet of the four
crystals. Within one material $B$ and $\ell_0$ do not track each other,
which is what the table is here to show. Each screw sector sums to zero at
$\Gamma$. For Se and Te a sector contains one acoustic and two optical
branches, and the acoustic branch is a circularly polarized uniform
translation carrying exactly $\hbar$, so the two optical values of $\ell_0$
sum to unity. For $\alpha$-SiO$_2$ and $\alpha$-HgS the same cancellation
involves more branches, with signs that the table does not display.

	The last column gives the difference
between the two branches of a doublet in $|\langle L_z\rangle|$. At $\Gamma$ the relation
$P_EL_zP_E=\hbar\ell_0\tau_z$ is exact, since $L_z$ and $\tau_0$ belong to
different representations of $D_3$. Away from $\Gamma$ the little group is
$C_3$, both become invariant, and a term $\hbar\ell_I(k_z)\tau_0$ odd in
$k_z$ is allowed, so that column is $2|\ell_I|$ at the edge of the fitting
window. We have checked that it is odd in $k_z$ to machine precision. It stays below $8$\% of
$\ell_0$ for the lowest doublet of each crystal and reaches $87$\% for the
$375$~cm$^{-1}$ doublet of $\alpha$-SiO$_2$, where $\ell_0$ itself is small. Values of $\ell_0$ below about $0.1$ should therefore be read
to one digit.

\begin{table}[h]
	\centering
	\caption{Every optical $E$ doublet along $\mathbf{k}\parallel c$.
	$\omega_0$ is the frequency at $\Gamma$, $B$ half the linear slope of the frequency difference,
	$b_1=2\omega_0B$ the splitting coefficient, and $\ell_0$ the magnitude of the angular momentum of either branch at $\Gamma$ in units of $\hbar$. The last column is the difference between the branches in $|\langle L_z\rangle|$ in units of $\hbar$, evaluated at the upper limit of the fitting range, $\zeta=0.04$. Magnitudes are given. The first entry of each
	crystal is the one quoted in Table~II of the main text. The doublets of each crystal are labelled $E^{(1)}$ onwards in order of increasing $\omega_0$.}
	\begin{tabular}{llrrrrr}
		\toprule
		& & $\omega_0$ (cm$^{-1}$) & $|B|$ (cm$^{-1}$\,\AA)
		& $|b_1|$ (cm$^{-2}$\,\AA) & $\ell_0$ & difference \\
		\midrule
		$\alpha$-SiO$_2$ & $E^{(1)}$ & $113.7$ & $77.9$ & $1.77\times10^{4}$
		   & $0.540$ & $0.014$ \\
		 & $E^{(2)}$ & $246.4$ & $87.3$ & $4.30\times10^{4}$
		   & $0.258$ & $0.007$ \\
		 & $E^{(3)}$ & $375.3$ & $66.9$ & $5.02\times10^{4}$
		   & $0.094$ & $0.081$ \\
		 & $E^{(4)}$ & $416.5$ & $71.6$ & $5.97\times10^{4}$
		   & $0.422$ & $0.013$ \\
		 & $E^{(5)}$ & $661.3$ & $238.4$ & $3.15\times10^{5}$
		   & $0.234$ & $0.030$ \\
		 & $E^{(6)}$ & $760.1$ & $8.0$ & $1.22\times10^{4}$
		   & $0.626$ & $0.083$ \\
		 & $E^{(7)}$ & $1041.0$ & $20.9$ & $4.35\times10^{4}$
		   & $0.101$ & $0.011$ \\
		 & $E^{(8)}$ & $1137.0$ & $101.6$ & $2.31\times10^{5}$
		   & $0.043$ & $0.001$ \\
		\midrule
		Se & $E^{(1)}$ & $127.8$ & $63.7$ & $1.63\times10^{4}$
		   & $0.806$ & $0.058$ \\
		   & $E^{(2)}$ & $216.0$ & $52.2$ & $2.26\times10^{4}$
		   & $0.194$ & $0.058$ \\
		\midrule
		Te & $E^{(1)}$ & $81.1$ & $32.7$ & $5.30\times10^{3}$
		   & $0.900$ & $0.032$ \\
		   & $E^{(2)}$ & $127.6$ & $32.3$ & $8.24\times10^{3}$
		   & $0.100$ & $0.033$ \\
		\midrule
		$\alpha$-HgS & $E^{(1)}$ & $43.4$ & $22.6$ & $1.96\times10^{3}$
		   & $0.887$ & $0.007$ \\
		             & $E^{(2)}$ & $84.5$ & $58.5$ & $9.87\times10^{3}$
		   & $0.086$ & $0.033$ \\
		             & $E^{(3)}$ & $116.4$ & $102.3$ & $2.38\times10^{4}$
		   & $0.189$ & $0.026$ \\
		             & $E^{(4)}$ & $265.7$ & $77.0$ & $4.09\times10^{4}$
		   & $0.215$ & $0.004$ \\
		             & $E^{(5)}$ & $323.1$ & $49.4$ & $3.19\times10^{4}$
		   & $0.204$ & $0.003$ \\
		\bottomrule
	\end{tabular}
	\label{tab:doublets}
\end{table}

\FloatBarrier

Although the splitting coefficient $b_1=2\omega_0B$ and the bulk
quantities $G_0$ and $G_u$ are all sensitive to structural handedness, they
characterize different aspects of the phonon dynamics. The former describes one
$E$ doublet near $\Gamma$, whereas the latter are thermally weighted sums over the
entire Brillouin zone.

\subsection{Comparison with Raman data and numerical stability}

Table~\ref{tab:raman} puts the $\alpha$-SiO$_2$ doublets at the wave
number reached in backscattering at $532$~nm and compares them with the
circularly polarized Raman measurement of Ref.~\cite{Oishi2024}. That measurement covers both enantiomers, and the
frequency differences interchange between them, which is the experimental
signature that the splitting reverses with the handedness of the
crystal. The measured peaks at $695$ and $128$~cm$^{-1}$ are tentatively associated with the computed doublets at $661$ and $114$~cm$^{-1}$, respectively. 
The circular configurations select only $E$ doublets. In the calculation $E^{(1)}$ is the only optical $E$ doublet below
$200$~cm$^{-1}$, the next being $E^{(2)}$ near $246$~cm$^{-1}$. This supports, but
does not establish, the assignment of the observed $128$~cm$^{-1}$ peak. The $695$~cm$^{-1}$ peak lies between two computed
doublets, at $4.8$ and $9.4$\%, so that assignment rests on
proximity alone.

\begin{table}[h]
	\centering
	\caption{Splitting of the $\alpha$-SiO$_2$ doublets in backscattering
	at $532$~nm, for which $k_z=4\pi n/\lambda=3.654\times10^{-3}$~\AA$^{-1}$
	with $n=1.5469$ taken from the same reference. The measured values are the relative differences reported
	in Ref.~\cite{Oishi2024} multiplied by the measured frequency. The associations between the measured peaks and the calculated doublets are tentative. Blank entries indicate that no experimental comparison is available.}
	\begin{tabular}{lrrrr}
		\toprule
		& $\omega_0$ (cm$^{-1}$) & $|\Delta\omega|$ (cm$^{-1}$)
		& measured peak (cm$^{-1}$) & measured $|\Delta\nu|$ (cm$^{-1}$) \\
		\midrule
 $E^{(1)}$ &		$113.7$ & $0.569$ & $128$ & $0.307$ \\
 $E^{(2)}$ &		$246.4$ & $0.638$ &  \\
 $E^{(3)}$ &		$375.3$ & $0.489$ &  \\
 $E^{(4)}$ &		$416.5$ & $0.523$ &  \\
 $E^{(5)}$ &		$661.3$ & $1.742$ & $695$ & $1.529$ \\
 $E^{(6)}$ &		$760.1$ & $0.059$ &  \\
 $E^{(7)}$ &		$1041.0$ & $0.153$ &  \\
 $E^{(8)}$ &		$1137.0$ & $0.742$ &  \\
		\bottomrule
	\end{tabular}
	\label{tab:raman}
\end{table}

Along $\mathbf{k}\parallel c$ the interpolated dynamical matrix is
\[
  D(0,0,\zeta)=\sum_{R_z}\Big[\sum_{R_\perp}C(R_\perp,R_z)\Big]\,
  e^{2\pi i\zeta R_z/c} ,
\]
so the dispersion on this line depends on the $q$-mesh only through
the number of divisions $N_c$ parallel to $c$, which is six for all four
crystals. We have verified this directly, since reducing the in-plane divisions of the mesh of Table~\ref{tab:params} to any divisor of their number leaves every digit of $\omega_0$ and of $B$ unchanged, and the subsets with $\mathbf{q}_\perp=0$ are themselves the complete meshes for $N_c=1$, $2$, $3$, and $6$.

	$B(N_c)$ was evaluated for every optical $E$ doublet. Most of them change little over those meshes, including the two compared with the Raman measurement, but two entries do not settle. The lowest doublet of Te changes by $18$\% and the second is not monotonic in $N_c$. These changes are the observed mesh dependence and not a bound on the error, since the increments for Te do not decelerate. These values test the $c$-axis mesh dependence of $B$, which is a
property of the $\mathbf{q}_\perp=0$ line alone. It is not a $c$-axis convergence
test of $G_0$ or $G_u$, which are sums over the whole zone and for which the
identity above does not apply.

The slopes above come from the interpolation, and for the two doublets of Te and the two of Se we have compared them with independently computed finite-$\mathbf{q}$ DFPT results. The dynamical matrix was computed directly at $\mathbf{q}=(0,0,\zeta)$ on the same cell and with the same settings as the production run, a route that involves neither interpolation nor an interpolation-based restoration of the long-range dipole term. The optical frequencies it gives at $\Gamma$ agree with the interpolated ones to $0.001$~cm$^{-1}$, which checks the two setups against each other at $\mathbf{q}=0$. Six values of $\zeta$ between $0.0025$ and $0.05$ were used. For Se the two routes agree to better than $1$\% on both doublets.

	For Te the direct slopes are $31.7$ and $42.5$~cm$^{-1}$\,\AA\ against interpolated values of $32.7$ and $32.3$~cm$^{-1}$\,\AA, so the interpolation is too large by $3$\% on $E^{(1)}$ and too small by $24$\% on $E^{(2)}$. Neither discrepancy is an artefact of the fitting window, since on $E^{(2)}$ the deviation is the same at every fitted $\zeta$ and on $E^{(1)}$ it varies only between $2.7$ and $3.6$\%. Switching off the restoration of the long-range dipole term moves the interpolated slope of $E^{(1)}$ by $9$\% but $E^{(2)}$ by $0.6$\% or less, so the large discrepancy does not come from that term either. The two Te slopes quoted in the text of the main paper are the direct ones, whereas Table~II of the main text and Table~\ref{tab:doublets} list the interpolated values. This comparison does not establish a general error bound for the interpolation.

\clearpage

\bibliography{chiral_phonons.bib,SM.bib}